\documentstyle[aaspp4,flushrt]{article}

\begin{document}

%%%%%%%%%%%%%%%%%%%%%%%%%%%%%%%%%%%%%%%%%%%%%%%%%%%%%%%%%%%%%%%%%%%%%%%%%%%%%
%
%                        TITLE PAGE AND ABSTRACT

\title{A Database of COBE-Normalized CDM Simulations (Abbreviated Version)}

\author{Hugo Martel\altaffilmark{1}
        and Richard Matzner\altaffilmark{2,3}}

\altaffiltext{1}{Department of Astronomy, University of Texas, Austin, 
                 TX 78712}
\altaffiltext{2}{Center for Relativity, University of Texas, Austin, TX 78712}
\altaffiltext{3}{Department of Physics, University of Texas, Austin, TX 78712}

\bigskip\bigskip

{\leftskip50pt\rightskip50pt

\noindent
{\bf This paper is an abbreviated version of our original manuscript.
We wrote this abbreviated version in order to meet the size limitations
imposed by the astro-ph archive. The original manuscript, which has
been submitted to The Astrophysical Journal, can be obtained by
contacting the authors.}

}
\bigskip\bigskip

\begin{abstract}
We have simulated the formation and evolution of large-scale structure
in the universe, for 68 different {\it COBE}-normalized
cosmological models. For each cosmological model, we have performed between 1
and 3 simulations, for a total of 160 simulations. This constitutes the
largest database of cosmological simulations ever assembled, and the
largest cosmological parameter space ever covered by such simulations. 
We are making this database available
to the astronomical community. We provide instructions for accessing
the database and for converting the data from computational units
to physical units.

The database includes Tilted Cold Dark Matter (TCDM) models,
Tilted Open Cold Dark Matter (TOCDM) models,
and Tilted $\Lambda$ Cold Dark Matter (T$\Lambda$CDM) models.
(For several simulations, the primordial exponent $n$ of the power spectrum
is near unity, hence these simulations can be considered as ``untilted.'')
The simulations cover a 4-dimensional cosmological
parameter phase space, the parameters being the present density parameter
$\Omega_0$, cosmological constant $\lambda_0$, and Hubble constant $H_0$,
and the rms density fluctuation $\sigma_8$ at scale $8h^{-1}\rm Mpc$. 
All simulations were performed
using a P$^3$M algorithm with $64^3$ particles on a
$128^3$ mesh, in a cubic volume of comoving size 128 Mpc. Each simulation
starts at a redshift of 24, and is carried up to the present.
More simulations will be added to the database in the future.

We have performed a limited amount of data reduction and analysis of
the final states of the simulations. We computed
the rms density fluctuation, the 2-point correlation function, the
velocity moments, and the properties of clusters. The details of these
calculations are presented in the full version of this paper. In this
abbreviated version, we present only the analysis of
the rms density fluctuation and two-point correlation function.
Our results are the following:

(1) The numerical value $\sigma_8^{\rm num}$ of the
rms density fluctuation differs from the value $\sigma_8^{\rm cont}$ obtained
by integrating the power spectrum at early times and extrapolating linearly
up the the present. This results from the combined effects of discreteness
in the numerical representation of
the power spectrum, the presence of a Gaussian factor in the 
initial conditions, and late-time nonlinear evolution.
The first of these three effects is
negligible. The second and third are comparable, and can both modify the
value of $\sigma_8$ by up to 10\%. Nonlinear effects, however, are important
only for models with $\sigma_8>0.6$, and can result in either an increase or
a decrease in $\sigma_8$. 

(2) The observed galaxy two-point correlation 
function is well reproduced by models with $\sigma_8\sim0.8$, nearly 
independently of the values of the other parameters, $\Omega_0$,
$\lambda_0$, and $H_0$. For models with 
$\sigma_8>0.8$, the correlation function is too large and its slope is 
too steep. For models with 
$\sigma_8<0.8$, the correlation function is too small, its slope is 
too shallow, and it often has a kink at separations
of order $1-3\,\rm Mpc$.

(3) At small separations, $r<1\,\rm Mpc$, 
the velocity moments indicate that small
clusters have reached virial equilibrium, while still accreting
matter from the field. The velocity moments depend essentially
upon $\Omega_0$ and $\sigma_8$, and not $\lambda_0$ and $H_0$.
The pairwise particle velocity dispersions are much 
larger than the observed pairwise galaxy velocity dispersion, for 
nearly all models. Velocity bias between galaxies and dark matter is needed 
to reconciled the simulations with observations.

(4) The cluster multiplicity function is decreasing for models with
$\sigma_8\sim0.3$. It has a horizontal plateau for models with
$\sigma_8$ in the range $0.4-0.9$. For models with $\sigma_8>0.9$,
it has a $\cup$ shape, which is probably a numerical artifact caused by the 
finite number of particles used in the simulations. For all models, 
clusters have densities
in the range 100--1000 times the mean background density, the
spin parameters $\lambda$ are in the range $0.008-0.2$, with the
median near 0.05, and about 2/3 of the clusters are prolate. Rotationally
supported disks do not form in these simulations.

\end{abstract}


\newpage
%%%%%%%%%%%%%%%%%%%%%%%%%%%%%%%%%%%%%%%%%%%%%%%%%%%%%%%%%%%%%%%%%%%%%%%%%%%%%
%
%                               TEXT

\def \kbf {{\bf k}}
\def \rbf {{\bf r}}
\def \xbf {{\bf x}}
\def \ybf {{\bf y}}
\def \zbf {{\bf z}}

\section{INTRODUCTION}

\subsection{Importance of Numerical Simulations in Cosmology}

Observations of the nearby universe reveal the existence of the large-scale
structure. The visible matter is clumped into galaxies, and these galaxies
are not distributed uniformly into space, but instead grouped into structures
such as clusters, filaments, and walls, separated by deep voids. 
Velocity structures (deviations from Hubble flow) are observed as well.
By contrast, observations of the Cosmic Microwave Background (CMB) reveal that
the universe was extremely uniform near the epoch of recombination. 
Hence, the present large-scale structure must result from an evolutionary
process that took place between recombination and the present. 
The most widely accepted scenario assumes that
the present large-scale structure originates from
the growth, by gravitational instability, of primordial density fluctuations
present in the early universe. Any fluctuation larger than the Jeans length
can grow by gravitational instability once the universe 
becomes matter-dominated. If the primordial density fluctuations originates
from a Gaussian random process (the usual assumption), then the 
primordial density field is entirely
described in terms of its power spectrum $P(k)$. The particular form of the power spectrum essentially depends upon the amount and nature of the
various components (baryonic matter, dark matter, cosmological constant,
and so on) present in the universe. If we assume a certain power spectrum, 
we can describe the primordial density field, and the formation and evolution
of large scale-structure in the universe becomes an initial value problem:
starting from the primordial density field, we can compute its evolution
using the laws of general relativity.
Unfortunately, this initial value problem is far too complex to be solved
analytically. We can simplify the problem by noticing that the largest 
structures observed in the universe are significantly smaller than the horizon.
This enables us to describe the evolution of the large-scale structure using 
Newtonian mechanics instead of general relativity (Peebles 1980,
Chapter 2). Even so, the general problem 
cannot be solved analytically. This leaves two possible
approaches: analytical approximations, or numerical simulations.

Two different kinds of analytical approximations have been
considered. The first one is based on the fact that
the initial fluctuations are small. We can expand the
equations describing the evolution of these fluctuations
in powers of the fluctuations, and solve them
using perturbation theory. This approach is extremely useful in describing 
the early evolution of the fluctuations, and has led to very
important results. However, it becomes inapplicable as soon as the fluctuations
become nonlinear. Such fluctuations still have to grow by a factor
of $10^2$ to reach the density of a cluster of galaxies, and $10^5$ to 
reach the density of a galaxy. Clearly, perturbation theory cannot be used
to describe the late stages of large-scale structure formation. 
The second analytical approach consists of
considering systems with a particular geometry
(see, for instance, Zel'dovich 1970; Peebles 1980, \S\S19--21;
Fillmore \& Goldreich 1984a, 1984b; Bertschinger 1985a, 1985b).
The most popular analytical models for large-scale structure formation are
the {\it Spherical Model}, which assumes spherical symmetry, and the
{\it Pancake Model}, which assumes planar symmetry. An important assumption
of some of 
these models is that the system considered is isolated. For instance,
the spherical model can describe the evolution of a self-gravitating spherical
overdensity, but we must
assume that any tidal influence from nearby structures
can be neglected, an assumption that might be valid at late times but
certainly not at early times.

These analytical approximations can therefore describe the universe at
early times or at late times, but not both. This problem can be solved by 
using mixed schemes, that combine various analytical approximations in a way
that allows an analytical description of the evolution of large-scale
structure at all epochs. The most important ones are the {\it Press-Schechter
Approximation} (Press \& Schechter 1979), 
which combines perturbation theory with the spherical model, and
the {\it Zel'dovich Approximation} (Zel'dovich 1970), 
which combines perturbation theory with 
the pancake model. 

The alternative consists of using numerical simulation. Unlike analytical
models, numerical simulations suffer from problems such as limited resolution
and numerical noise. Also, simulations provide very little insight into
the physical processes taking place, compared to analytical models. 
However, numerical simulations can
describe the evolution of the large-scale structure entirely, from the
initial conditions all the way to the present, without making any 
approximation or imposing any restriction on the geometry of the system.
Cosmological N-body simulations have played a central role in the study of
the formation and evolution of large-scale structure in the universe
during most of the last two decades. These simulations have contributed
the improve our understanding of the physical process of gravitational
instability that leads to structure formation, have enable us to 
conceive and test various cosmological scenarios, 
and have produced simulated universes that can directly be compared with 
observations (Efstathiou \& Eastwood 1981; Centrella \& Melott 1983;
Klypin \& Shandarin 1983; Miller 1983; Shapiro, Struck-Marcell, \& Melott 1983;
White, Frenk, \& Davis 1983; Davis et al. 1985; Efstathiou et al. 1985;
Barnes \& Hut 1986, 1989; Evrard 1986, 1987; Melott 1986; White et al. 1987a,
1987b; Frenk et al. 1988; Gramann 1988; Carlberg \& Couchman 1989;
Villumsen 1989; West, Oemler, \& Dekel 1989; Couchman 1991; 
Fukushige et al. 1991; Hernquist, Bouchet, \& Suto 1991; Martel 1991a; 
Moutarde et al. 1991; West, Villumsen, \& Dekel 1991; 
Bouchet \& Hernquist 1992; Fry, Melott, \& Shandarin 1992; Park et al. 1992;
Bahcall, Cen, \& Gramann, 1993; Gramann, Cen, \& Bahcall 1993;
Melott \& Shandarin 1993; Babul et al 1994; Pen 1995; 
Colombi, Bouchet, \& Hernquist 1996; Moore, Katz, \& Lake 1996: 
Yess \& Shandarin 1996; Klypin, Nolthenius, \& Primack 1997; 
Kravtsov, Klypin, \& Khokhlov 1997; Navarro, Frenk, \& White 1997; 
Gross et al. 1998; Thomas et al. 1998).\footnote{Cosmological numerical 
simulations have been used intensively since 1981, and their results 
have appeared in hundreds of publications, so this list is necessarily 
incomplete. We decided to include only the key publications by each 
research group. We also excluded one- and
two-dimensional simulations (for brevity), and
simulations with hydrodynamics, which involve the next
generation of numerical algorithms.}

\subsection{The Standard Model}

Theoretical developments in particle theory and early-universe physics,
combined with numerical simulations and observations of the large-scale
structure of the universe, led to the emergence during the 1980's of
what became known as the {\it Standard Cosmological Model}. The
inflationary scenario (Guth 1981; Linde 1982; Albretch \& Steinhardt 1982)
requires that the universe is spatially flat to a great accuracy. 
In a matter-dominated universe, in the absence of any exotic components
such as a nonzero cosmological constant $\Lambda$, this requires that the mean
density of the universe is equal to its critical density, or, alternatively,
\begin{equation}
\Omega_0=1\,,
\end{equation}

\noindent where $\Omega_0\equiv8\pi G\bar\rho_0/3H_0^2$ is the density 
parameter, $\bar\rho_0$ is the mean density of the universe, and $H_0$
is the Hubble constant (throughout
this paper, we use subscripts 0 to designate present values). 
In this scenario, the
large-scale structure of the universe that we observe today results from
the growth, by gravitational instability, of small density perturbations
present at recombination, which originate from quantum processes in the 
early universe. There are two difficulties with this scenario.
First, there is strong evidence that the amount of ``ordinary matter'' in the 
universe is insufficient to satisfy equation (1). Primordial nucleosynthesis
provides stringent upper limit to the baryonic content of the universe,
and shows that the present baryon contribution to the density parameter,
$\Omega_{{\rm B}0}$, is less than $0.026h^{-2}$, 
where $h$ is the Hubble constant in units of $100\rm\,km\,s^{-1}Mpc^{-1}$ 
(Krauss \& Kernan 1995; Copi, Schramm, \& Turner 1995; Krauss 1998). 
Furthermore, dynamical studies
of rich clusters of galaxies show that the contribution to
the density parameter of the matter that clusters at 
that scale is $\Omega_{\rm clusters}=0.2\pm0.1$
(Gott et al. 1974; Carlberg et al. 1996; Lin et al. 1996).
Second, observations of the temperature fluctuations
in the Cosmic Microwave Background (CMB) provide an upper
limit to the density fluctuations at recombination, of order 
$\delta\rho/\rho\sim10^{-5}$. Such fluctuations would grow by gravitational
instability, to reach an amplitude of order $10^{-2}$ by the present,
clearly insufficient to explain the
origin of large-scale structure and galaxies.

These two difficulties were solved by postulating the existence of a
component known as {\it dark matter}. This dark matter,
which can be detected only through its gravitational
influence, makes up the difference between
the amount of matter required to satisfy equation~(1) and the amount of
matter that is observed or indirectly measured. We can then reconcile the
dynamical estimates of $\Omega_0$ with equation~(1) by assuming that the
dark matter is distributed more smoothly than the 
luminous matter that makes up
galaxies and clusters, an idea known as {\it biasing} (Kaiser 1984). 
However, since 
the dynamical estimates of $\Omega_0$ exceed the limit imposed 
by primordial nucleosynthesis, some amount of dark matter must be clustered
on galactic and cluster scales, even though the bulk of the dark matter is
smoothly distributed into space. Also, density fluctuations in the dark
matter start growing when the universe becomes matter-dominated, and 
by the time recombination occurs, the dark matter fluctuations have already
grown by a factor of $\sim20h^2/\Omega_{\rm B0}$
(Kolb \& Turner 1990, \S9.5). These fluctuations
provide potential wells into which the baryons
fall soon after recombination.
This enables us to reconcile the CMB measurements,
--- which are sensitive to the fluctuations in the baryonic matter, but not 
the dark matter --- with the existence of galaxies and large-scale structure,
Many candidates for dark matter
have been suggested, several of them emerging from particle theory.
These various forms of dark matter are usually classified as
Hot Dark Matter (HDM) and Cold Dark Matter (CDM).

Cosmological numerical simulations performed in the 1980's
have shown that an $\Omega_0=1$ universe in which most
matter is in the form of Cold Dark Matter, and the distribution of
luminous matter is biased relative to the distribution of dark matter,
can successfully reproduce all
observations of large-scale structure that were then known 
(see, e.g. Davis et al. 1985), while satisfying the constraints imposed
by the inflationary scenario, primordial nucleosynthesis, and
the CMB (unlike HDM, which as difficulties explaining structure
formation at galactic scales). 
These simulations played a central role in establishing the 
$\Omega_0=1$, biased CDM model as the Standard Cosmological Model.

\subsection{Non-Standard Models}

The Standard Model, which has been hailed by many as the ``final answer''
to the problem of structure formation and evolution in the universe,
ran into serious problems during the 1990's.
In this subsection, we briefly review these problems.

\subsubsection{The Age Problem}

In a flat, matter-dominated universe, the age of the universe $t_0$
is 2/3 of the Hubble time, that is $t_0=2/3H_0=6.52\times10^9h^{-1}\rm years$.
For $h$ in the range $0.5-1$, this
corresponds to an age in the range $6.52-13.04\times10^9\rm\,years$.
Measurements of the ages of globular clusters indicate that the oldest
clusters are certainly older than $9.5\times10^9\rm\,years$,
and most likely in the range $11-13\times10^9\rm\,years$
(Jimenez et al. 1996; Chaboyer 1998; Chaboyer et al. 1998; Jimenez 1998).
These measurements are only marginally consistent with the standard model, 
in the sense that they require a Hubble constant near its smallest
possible value, $h\sim0.5$. However, recent observations have significantly
reduced the range of plausible values for the Hubble constant,
showing that $h$ is likely to be in the range
$0.65-0.75$ (Freedman, W. L. 1998, and references therein).
In the standard model, this corresponds to an age in the range
$8.69-10.03\times10^9\rm\,years$. The upper end of this range is still
compatible with the measured ages of globular clusters, but just barely. 

\subsubsection{Large-Scale Structure}

Until 1992, the amplitude of the primordial density fluctuation
power spectrum was unknown. We were free to tune this amplitude in
order to reproduce the correct amount of galaxy clustering observed today.
The latter is usually characterized by the 
rms density fluctuation $\sigma_8$ at a scale of $8h^{-1}\rm Mpc$. 
Observations of clusters of galaxies show that $\sigma_8\approx0.6$ for 
the standard model (Viana \& Liddle 1996, see eq.~[42] below). The
discovery by the {\it COBE} DMR experiment of degree-scale fluctuations
in the CMB temperature (Smoot et al. 1992)
has eliminated this freedom, by fixing the amplitude
power spectrum. A {\it COBE}-normalized Standard Model produces too
much structure at cluster scales (Barlett \& Silk 1993).
The resulting value of $\sigma_8$ is
$1.22$ for $h=0.5$ (Bunn \& White 1997), too large
by a factor of 2, and becomes even larger for larger $h$. 
By combining the {\it COBE} result with observations of the present
large-scale structure, we obtain a constraint on the quantity $\Omega_0h$,
which is
\begin{equation}
0.2\leq\Omega_0h\leq0.3
\end{equation}

\noindent (Peacock \& Dodds 1994). Since $h$ is certainly larger than 0.5,
this implies $\Omega_0<0.6$.

\subsubsection{The Baryon Catastrophe}

In the Standard Model, the baryon fraction of the universe is small.
Primordial nucleosynthesis imposes the constraint that
$\Omega_{\rm B0}h^2<0.026$
(Krauss \& Kernan 1995; Copi, Schramm,
\& Turner 1995; Krauss 1998). For $h=0.65$,
this corresponds to $\Omega_{\rm B0}=0.061$. Hence, if $\Omega_0=1$,
at most 6\% of the matter if the universe is composed of baryons,
the rest being dark matter. However, observations of X-ray clusters
reveal that the baryon fraction in these clusters is $\sim0.1h^{-1.5}$,
or 19\% for $h=0.65$ (Briel, Henry, \& Boringer 1992). 
Hence, X-ray clusters contain a large excess of baryons relative to dark
matter compared with the average values in the universe, a situation
often referred to as {\it the baryon catastrophe}. 
This problem could be 
solved if we can think of a physical process that would concentrate the 
baryons inside clusters, creating a bias relative to the dark matter.
However, no such physical process is known. It is much simplier to assume
that the density parameter is less than unity. In this case, the universal 
baryon fraction is not $\Omega_{\rm B0}$, but $\Omega_{\rm B0}/\Omega_0$.
This baryon fraction is $\sim0.1h^{-1.5}$ according to
observations of X-ray clusters, and smaller than $0.026/\Omega_0h^2$ according
to primordial nucleosynthesis. Combining these two results, we get
\begin{equation}
\Omega_0h^{1/2}<0.26\,.
\end{equation}

\noindent This rules out the Standard Model (unless $h<0.07$ !). For $h>0.5$,
this limit becomes $\Omega_0<0.37$.

\subsubsection{Evolution of Cluster Abundance}

In the Standard Model, the density parameter $\Omega$ 
is unity at all times, and density fluctuations can grow by gravitational
instability all the way to the present. 
In other models, density fluctuations can grow at
early times, when $\Omega$ is near unity.
But eventually, $\Omega$ drops significantly below unity,
and the density fluctuations ``freezes-out.'' Hence, in a model with 
$\Omega_0<1$, the present abundance of clusters should be comparable to
the abundance immediately after freeze-out, since not much growth has
taken place since then. Conversely, in the Standard Model, the present
abundance of clusters should be larger than the past one, since the growth
of density fluctuations never freezes out.
Bahcall, Fan, \& Cen (1997) and  Bahcall \& Fan (1998) have determined
the mass of three massive distant clusters, located at redshift $z>0.5$,
and showed that in a $\Omega_0$ universe, there should be only $\sim10^{-3}$
such clusters at $z>0.5$. They conclude that the density parameter
is in the range
\begin{equation}
0.1<\Omega_0<0.35
\end{equation}

\noindent (Bahcall 1999).

\subsubsection{Distant Type I Supernovae}

The relationship between the luminosity distance $D_L$ and the redshift 
is model-dependent. If standard candles can be observed at cosmological
distances, then the $D_L(z)$ relationship can be inferred, and limits
can be placed on the value of the cosmological parameters. This method was
recently applied to samples of distant (``High-$z$'') Type I supernovae,
by two independent research teams (Garnevich et al. 1998, and
references therein; Perlmutter et al. 1998, and references
therein). Applied to models with a nonzero cosmological constant $\Lambda$,
their observations provide severe constraints in the $\Omega_0-\lambda_0$
phase space (where $\lambda_0\equiv\Lambda/3H_0^2$). Not only the
Standard Model is excluded with a high degree of confidence, but open
models ($\Omega_0<1$), {\it without} a cosmological constant are also excluded,
unless $\Omega_0$ is very small. Observations of the CMB (White 1998;
Tegmark et al. 1998) provide a different constraint, that does not rule
out the Standard Model. However, combining the CMB and Type I supernovae
observations
leads to separate determinations of $\Omega_0$ and $\lambda_0$. The
preferred values are $\Omega_0\sim0.3$ and $\lambda_0\sim0.7$.

\subsubsection{Anthropic Considerations}

In models such as chaotic inflation,
in which the observed big bang is just one of an infinite number of expanding
regions in each of which the fundamental property takes a different value
(Linde 1986, 1987, 1988), and models in which a state vector is derived for 
the universe which is a superposition of terms with different values of 
the fundamental property (e.g. Hawking 1983, 1984; Coleman 1988), the 
probability of observing any particular values of the cosmological parameters 
is conditioned by the existence of observers in those ``subuniverses'' in 
which the parameters take these values (Efstathiou 1995; Vilenkin 1995; 
Weinberg 1996; Martel, Shapiro, \& Weinberg 1998).
This probability is proportional to the fraction of matter which is
destined to condense out of the background into mass concentrations
large enough to form observers.
Using this approach, Martel et al. (1998) calculated the relative likelihood
of observing any given value of the cosmological constant
$\Lambda$ within the context of the
flat CDM model normalized to {\it COBE}, and found that small
but {\it finite} value of the cosmological
constant, in the range suggested by observations, are favored
over the value $\Lambda=0$. Garriga, Tanaka, \& Vilenkin (1998)
have performed a similar analysis, but applied the the density parameter,
and found that intermediate values of $\Omega_0$ are more
likely to be observed than values near 0 or near 1. Anthropic arguments
do not favor the values that the parameters take in the
Standard Model, $\Omega_0=1$ and $\lambda_0=0$.

\subsubsection{Alternatives to the Standard Model}

The problems listed above strongly argue against the Standard Model, and
forces us to consider alternatives. The age problem, large-scale structure 
problem, and the baryon catastrophe can all be solved by considering Open
CDM, or OCDM models, in which the density parameter $\Omega_0<1$. However,
such models do not satisfy the flatness requirement of inflation. 
CDM models with a nonzero cosmological constant $\lambda_0$ equal
to $1-\Omega_0$, known as $\Lambda$CDM models, satisfy this flatness
requirement, and the addition of the cosmological constant improves the
age and large-scale structure problems, while providing a better
agreement to the supernovae data. 
Recently, several authors have shown that it is possible to reconcile the 
inflationary scenario with an open universe, thus eliminating the flatness
requirement (Ratra \& Peebles 1994;
Bucher, Goldhaber, \& Turok 1995; Yamamoto, Sasaki, \& Tanaka
1995; Linde 1995; Linde \& Mezhlumian 1995). This not only supports open,
matter-dominated models, but also allows for the possibility of 
an open universe with $\lambda_0\neq0$ and $\Omega_0+\lambda_0<1$.

The large-scale structure problem
can also be solved by introducing a ``tilt'' in the primordial power spectrum.
In this Tilted CDM, or TCDM model, the primordial power spectrum $P(k)$ at 
large scales does no have the Harrison-Zel'dovich form $P(k)\propto k$,
but instead varies as $P(k)\propto k^n$, where the primordial exponent
$n$ can differ from unity. The universe might also
contain a mixture of two different forms of dark matter, one cold and one hot,
a model known as CHDM. Finally, the universe might contain a smooth
component whose pressure $p$ and density $\rho$ are related by
an equation of state $p=w\rho$, a concept known as ``quintessence'' 
(Caldwell, Dave, \& Steinhardt 1998; see also Fry 1985; Charlton \& Turner 
1987; Silveira \& Waga 1994; Martel 1995; Martel \& Shapiro 1998).
The cosmological constant
is a particular form of quintessence, corresponding to $w=-1$; 
other forms have been suggested, such as
domain walls, textures, or strings.

The Standard Model had no free parameters. The values of $\Omega_0$ and
$\lambda_0$ were fixed at 1 and 0, respectively. The value of $h$ had
to be close to $0.5$ in order to avoid conflicts with ages of globular
clusters, and the primordial power spectrum was assumed to be a CDM
spectrum with no tilt. With the emergence of alternative models, there are
now many free parameters. The density parameters $\Omega_0$
no longer has to be unity. The cosmological constant $\lambda_0$ can be
nonzero and, if open inflation is correct, a nonzero $\lambda_0$ does not
have to be equal to $1-\Omega_0$. The Hubble constant can vary over
a certain range without conflicting with observations, and the slope $n$
of the primordial power spectrum does not have to be 1. In models such
as CHDM and quintessence models, there are additional parameters: the
contribution of each component to the density parameter, and in the case
of quintessence models, the coefficient $w$ appearing in
the equation of state.

During the 1980's, N-body simulations have played a central role in
establishing the Standard Model, and then went more or less into
hibernation, as efforts were invested into adding more physics to the
original algorithms (hydrodynamics in particular). 
The emergence of alternative cosmological models has lead to
a renewal of interest in N-body simulations. Such numerical simulations are 
essential for testing cosmological models against observations. Furthermore,
they are useful from a theoretical viewpoint, since they can reveal
how each cosmological parameter affects the process of large-scale structure
formation.

\subsection{The Need for a Database}

Numerical methods such as the {\it Particle-Mesh}
algorithm (PM) and the {\it Particle-Particle/Particle-Mesh} algorithm
($\rm P^3M$) are well documented. Details of the algorithms can be found
in textbooks (e.g. Hockney \& Eastwood 1981) and papers
(e.g. Efstathiou et al. 1985). Hence, any researcher can easily access all
the information and knowledge necessary to develop such algorithms.
However, the effort required to develop, test, and optimize 
a PM or $\rm P^3M$ algorithm from scratch can be quite substantial, and
can be regarded as a waste of effort, since it essentially amounts to
``reinventing the wheel.'' Also, performing simulations with large number
of particles can demand a substantial 
investment in resources such as computer time,
which is also wasteful if these simulations, or similar ones, have already
been performed by other researchers. Consequently, it is a common practice 
among researchers to share either their programs or the results of their
simulations.

Klypin \& Holtzman (1997) have combined into a single package their version 
of the PM algorithm and programs for generating initial conditions
and analyzing the results. This package
has been made available to the astronomical community, and can
can be downloaded from a world-wide-web site.
This allows other researchers interested
in performing cosmological numerical simulations to ``get started'' 
immediately, without having to develop and test any computer program. However,
installing and running these programs might pose some difficulties depending
upon the kind of computer resources available to the user.

We use a different approach. Instead of making our programs available
to the astronomical community (something we might do eventually),
it is the results of the simulations themselves that we are making available.
We performed a very large number of numerical simulations, a total of
160, for 68 different cosmological models. This constitutes by far
the largest database of cosmological simulations ever assembled, and it is
still growing as more simulations are being performed. We are making 
this database available to the astronomical community (see \S3.4 below).
This approach is complementary to the one used by Klypin \& Holtzman.
By providing the results of the simulations, we eliminate the need for 
researchers to perform themselves these simulations, and the same simulations 
can be used by many different researchers. 
However, someone might be interested in simulating
a cosmological model which is not included in the database, in which case
the algorithm of Klypin \& Holtzman can be used. Alternatively, we can,
upon request, perform additional simulations and include them in the database. 
An interesting question is whether the results of 
simulations from the database can be analyzed using Klypin \& Holtzman
programs. In principle, this should be possible. The output files in
the database are not written in the same format as the ones produced by 
Klypin \& Holtzman's PM code, but it is fairly trivial to write a program
that ``translate'' files from one format to another.

There is an important difference that must be pointed out. The program
of Klypin \& Holtzman is based on the PM algorithm, while the simulations
in the database were performed using a $\rm P^3M$ algorithm.
For a same number of particles, the $\rm P^3M$ algorithm 
has a length resolution
superior to the one of the PM algorithm by a factor of order 6 (depending
upon the particular choice of smoothing length). However, since the PM
algorithm is significantly faster than the $\rm P^3M$ algorithm, it is possible
to make up for the lack of resolution of the PM code by simply using
more particles. We used $64^3$ particles in all simulations.\footnote{We intend
to add simulations with $128^3$ particles to the database in a near future.}
These have the same length resolutions as PM calculations with $\sim384^3$
particles (such as the ones performed by Gross et al. [1998]).

As mentioned above, there are numerous alternatives to the Standard Model.
In this paper we consider CDM models in which the only components are
ordinary matter (dark and baryonic) and possibly a nonzero cosmological 
constant (thus excluding CHDM models, and generic quintessence models). 
We consider the three cases $\Omega_0=1$, $\lambda_0=0$ 
(the Einstein-de~Sitter model), $\Omega_0<1$, $\lambda_0=0$, and
$\Omega_0+\lambda_0=1$. We also allow the primordial
power spectrum to have a tilt. These models are usually referred
to as Tilted CDM (TCDM), Tilted, Open CDM (TOCDM), and Tilted,
Lambda CDM (T$\Lambda$CDM).

\section{THE NUMERICAL SIMULATIONS} 

\subsection{The Algorithm}

All simulations presented in this paper
were done using the P$^3$M algorithm (Hockney \&~Eastwood 1981;
Efstathiou et al. 1985). 
The computational volume is a cubic box of comoving size $L_{\rm box}$ and
comoving volume $V_{\rm box}=L_{\rm box}^3$ with 
triply periodic boundary conditions, expanding with Hubble flow.
The matter distribution inside the computational volume
is represented by $N$ equal-mass particles.
The forces on particles are computed by solving Poisson's equation on a 
cubic grid using a Fast Fourier Transform method.
The resulting force field represents the Newtonian interaction
between particles down to a separation of a few mesh spacings. At shorter
distances the computed force is significantly smaller than the
physical force. To increase
the dynamical range of the code, the force at short distance 
is corrected by direct 
summation over pairs of particles separated by less than some 
cutoff distance~$r_e$. With the addition of this so-called
{\it short-range correction}, the code accurately reproduces the Newtonian
interaction down to the softening length~$\epsilon$, which is a fraction
of the grid spacing. The system is evolved forward in time using a 
second order Runge-Kutta time-integration scheme with a variable time step.

Our particular version of the $\rm P^3M$ algorithm uses 
{\it supercomoving variables} (Martel \& Shapiro 1998; see also
Shandarin 1980). In these variables,
the position $\bf\tilde r$, peculiar velocity $\bf\tilde v$, 
time $\tilde t$, density $\tilde\rho$, and peculiar gravitational potential 
$\tilde\phi$ are related to their Eulerian counterparts by
\begin{eqnarray}
{\bf\tilde r}&=&{{\bf r}\over ar_*}\,,\\
{\bf\tilde v}&=&{a{\bf v}t_*\over r_*}\,,\\
{d\tilde t}&=&{dt\over a^2t_*}\,,\\
{\tilde \rho}&=&{a^3\rho\over\rho_*}\,,\\
{\tilde \phi}&=&{a^2\phi t_*^2\over r_*^2}\,,
\end{eqnarray}

\noindent
where 
\begin{eqnarray}
\rho_*&=&\bar\rho_0={3H_0^2\Omega_0\over8\pi G}\,,\\
t_*&=&{2\over H_0(\Omega_0a_0^3)^{1/2}}\,.
\end{eqnarray}
 
\noindent In these equations, $a(t)$ is
the Friedmann-Robertson-Walker scale factor, $a_0$ is its present value,
and $r_*$, is a free parameter whose value is
chosen according to the characteristic length scale of the problem.
These variables are similar to the more standard comoving
variables in many respects. In particular, equations (5) and (8)
imply that a volume expanding with Hubble flow remains fixed in
supercomoving variables, and that the mean density inside
that volume remains constant. The main difference is
in the change of time variable given by equation~(7). In supercomoving
coordinates, the time $\tilde t$ is negative, and equal to $-\infty$ at
the big bang. In an Einstein-de~Sitter model, $\tilde t=-1$ at present.
This change of time variable has the virtue of eliminating the
cosmological drag term in the momentum equation.

In all simulations, we set $r_*=L_{\rm box}/a_0$. Equation~(5) then
implies that the box size in supercomoving variables is unity at all times.

The time-evolution of the scale factor $a(t)$ is governed by
the Friedmann equation. For universes composed of ordinary, 
nonrelativistic matter and a nonzero cosmological constant $\lambda_0$, the
Friedmann equation takes the form
\begin{equation}
\biggl({1\over a}{da\over dt}\biggr)^2=H(t)^2
=H_0^2\Biggl[(1-\Omega_0-\lambda_0)\biggl({a\over a_0}\biggr)^{-2}
+\Omega_0\biggl({a\over a_0}\biggr)^{-3}
+\lambda_0\Biggr]\,.
\end{equation}

\noindent
In supercomoving variables, there is a precise normalization for the
scale factor, which depends upon the particular cosmological model. For
the models considered in this paper, the solution of the Friedmann
equation and the present value of the scale factor are the following:

\noindent (a) Einstein-de~Sitter model ($\Omega_0=1$, $\lambda_0=0$)
\begin{equation}
a=\tilde t^{-2}\,,\qquad a_0=1\,.
\end{equation}

\noindent (b) Open models ($\Omega_0<1$, $\lambda_0=0$)
\begin{equation}
a=(\tilde t^{\,2}-1)^{-1}\,,\qquad a_0=(1-\Omega_0)/\Omega_0\,.
\end{equation}

\noindent (c) Flat models with nonzero cosmological constant
($\Omega_0+\lambda_0=1$)
\begin{equation}
\tilde t={1\over2}\int_1^a {dy\over y^{3/2}(1+y^3)^{1/2}}\,,
\qquad a_0=\biggl({\lambda_0\over\Omega_0}\biggr)^{1/3}\,.
\end{equation}

\noindent Notice that the solutions for $a(\tilde t\,)$ 
do not depend explicitly
upon the cosmological parameters, which are absorbed in the definition of
$a_0$. Hence, for all models included in the database, there are only 3
different solutions of the Friedmann equation. This is one of the most
useful properties of supercomoving variables.
For simplicity, we shall drop the tilde notation for
supercomoving variables in the remainder of this paper, except in \S3.3.

\subsection{The Power Spectrum}

For all simulations presented in this paper, 
we use the Cold Dark Matter (CDM) power spectrum of
Bardeen et al. (1986), with the normalization of Bunn \&~White (1997).
The power spectrum at redshift $z$ is given by
\begin{equation}
P(k,z)=2\pi^2\biggl({c\over H_0}\biggr)^{3+n}\delta_H^2{\cal L}^{-2}(z,0)
k^nT_{\rm CDM}^2(k)\,,
\end{equation}

\noindent where $c$ is the speed of light,
${\cal L}(z,0)\equiv\delta_+(0)/\delta_+(z)$  
is the linear growth factor between redshift $z$ and the present, and
$\delta_+$ is the linear growing mode (see eqs.~[35]--[39] below),
$n$ is the tilt,
and $T_{\rm CDM}$ is the transfer function, given by
\begin{equation}
T_{\rm CDM}(q)={\ln(1+2.34q)\over2.34q}\big[1+3.89q+(16.1q)^2+(5.46q)^3
+(6.71q)^4\big]^{-1/4}
\end{equation}

\noindent (Bardeen et al. 1986), with $q$ is defined by
\begin{eqnarray}
q&=&\biggl({k\over{\rm Mpc}^{-1}}\biggr)\alpha^{-1/2}(\Omega_0h^2)^{-1}
\Theta_{2.7}^2\,,\\
\alpha&=&a_1^{-\Omega_{\rm B0}/\Omega_0}a_2^{-(\Omega_{\rm B0}/\Omega_0)^3}\,,\\
a_1&=&(46.9\Omega_0h^2)^{0.670}\big[1+(32.1\Omega_0h^2)^{-0.532}\big]\,,\\
a_2&=&(12.0\Omega_0h^2)^{0.424}\big[1+(45.0\Omega_0h^2)^{-0.582}\big]
\end{eqnarray}

\noindent (Hu \& Sugiyama 1996, eqs.~[D-28] and [E-12]), where $\Theta_{2.7}$ 
is the temperature of the cosmic microwave background in units of 2.7K,
and $\delta_H$ is the density perturbation at horizon crossing
(Liddle \& Lyth 1993). Fits for $\delta_H$ are given by Bunn \& White (1997), 
as follows,
\begin{equation}
10^5\delta_H=\cases{
1.95\Omega_0^{-0.35-0.19\ln\Omega_0-0.17\tilde n}e^{-(\tilde n+0.14\tilde n^2)}
\,,& $\lambda_0=0$;\cr
1.94\Omega_0^{-0.785-0.05\ln\Omega_0}e^{-(0.95\tilde n+0.169\tilde n^2)}\,,
& $\lambda_0=1-\Omega_0$;\cr
%2.422+(-1.166+3.392\lambda_0)e^{\Omega_0}+(0.800-2.267\Omega_0)e^{\lambda_0}
%&\cr\qquad
%+3.780\Omega_0+0.561\lambda_0+0.487\Omega_0^2-8.568\Omega_0\lambda_0
%+1.080\lambda_0\,,&$n=1$\,;\cr
}
\end{equation}

\noindent where $\tilde n\equiv n-1$.

\subsection{Setting up Initial Conditions}

We assume that the initial fluctuations originate from a Gaussian random
process. The initial density contrast can then be represented as a superposition
of plane waves with random phases, and amplitudes related to the
power spectrum $P(k)$, where $\kbf$ is the wavenumber, and
$k=|{\kbf}|$. In an infinite universe, all values of $\kbf$ are
allowed. The power spectrum is therefore continuous, and the
number of modes (that is, plane waves) present in the initial density
contrast is infinite. The simulations, however, are performed inside
a finite comoving cubic volume $V_{\rm box}=L_{\rm box}^3$ with periodic
boundary conditions. This periodicity implies that only modes with wavenumbers
$\kbf=(k_x,k_y,k_z)=(n_x,n_y,n_z)k_0$, where $n_x$, $n_y$, $n_z$
are integers and $k_0\equiv2\pi/L_{\rm box}$ is the fundamental wavenumber,
can be present in the simulated initial conditions. Furthermore, since the
initial conditions are represented by particles, the components
$k_x$, $k_y$, $k_z$ of the wavenumber cannot exceed the nyquist frequency
$k_{\rm nyq}=N^{1/3}k_0/2$, where $N$ is the number of particles in
the computational volume, and $N^{1/3}$ is the number of particles along
one dimension. Modes with higher wavenumber cannot be represented because
of undersampling. Hence we are faced with the task of representing 
continuous initial conditions using a discrete sample of plane waves.
This key aspect of any numerical cosmological simulation is, surprisingly,
seldom discussed in the literature. Here we present a detailed description.

In a periodic universe with comoving cubic volume
$V_{\rm box}=L_{\rm box}^3$, the density contrast $\delta$ can be
decomposed into a sum of plane waves,
\begin{equation}
\delta(\rbf)=\sum_\kbf\delta_\kbf^{\rm disc}e^{-i\kbf\cdot\rbf}\,,
\end{equation}

\noindent where $\rbf$ is the comoving, or supercomoving, 
position, and $\delta_{\kbf}^{\rm disc}$
is the amplitude of the mode with wavenumber $\kbf$. The superscript ``disc''
stands for ``discrete.''
The real universe is of course not periodic, in which case all values
of $\kbf$ are allowed. To convert equation (23) from the
discrete limit to the continuous limit, consider 
first any function $f(\kbf)$ that
is summed over all possible values of $\kbf$. In the discrete limit, we have
\begin{equation}
\sum_\kbf f_\kbf^{\rm disc}=\sum_{\rm all\>V.E.} f_\kbf^{\rm disc}
={1\over k_0^3}\sum_{\rm all\>V.E.} f_\kbf^{\rm disc}k_0^3
={V_{\rm box}\over(2\pi)^3}\sum_{\rm all\>V.E.} f_\kbf^{\rm disc}
\int_{\rm V.E.}d^3k\,,
\end{equation}

\noindent where ``V.E.'' represents a {\it volume element} in $\kbf$-space,
which is a cube of volume $k_0^3=V_{\rm box}/(2\pi)^3$ 
centered at $\kbf$. Assuming that the
function $f$ does not vary significantly over one volume element, we can pull
it inside the integral,
\begin{equation}
\sum_\kbf f_\kbf^{\rm disc}\approx{V_{\rm box}\over(2\pi)^3}
\sum_{\rm all\>V.E.}\int_{\rm V.E.}f_\kbf^{\rm disc}d^3k\,.
\end{equation}

\noindent Of course, integrating over the volume element, and
then summing over all volume elements, is effectively like integrating
over all $\kbf$-space, so equation~(25) reduces to
\begin{equation}
\sum_\kbf f_\kbf^{\rm disc}\approx{V_{\rm box}\over(2\pi)^3}
\int f_\kbf^{\rm disc}d^3k=\int f_\kbf^{\rm cont}d^3k\,,
\end{equation}

\noindent where the superscript ``cont'' stands for
``continuous.'' The continuous and discrete functions are related by
\begin{equation}
f_\kbf^{\rm cont}={V_{\rm box}\over(2\pi)^3}f_\kbf^{\rm disc}\,.
\end{equation}

\noindent Using these formulae, we can rewrite equation~(23) as
\begin{equation}
\delta(\rbf)=\int d^3k\,\delta_\kbf^{\rm cont}e^{-i\kbf\cdot\rbf}\,,
\end{equation}

\noindent where
\begin{equation}
\delta_\kbf^{\rm cont}={V_{\rm box}\over(2\pi)^3}
\delta_\kbf^{\rm disc}\,.
\end{equation}

To find the relationships between $\delta_\kbf^{\rm disc}$,
$\delta_\kbf^{\rm cont}$, and the power spectrum, consider the rms 
density fluctuation $\sigma_x$ at some particular scale $x$. 
This quantity is given by
\begin{equation}
\sigma_x^2={V_{\rm box}\over(2\pi)^3}
\int d^3k|\delta_\kbf^{\rm disc}|^2W(kx)\,.
\end{equation}

\noindent where $W$ is a window function. We present the derivation of this 
result in Appendix~A. In the continuous limit, $\sigma_x$ is related to
the power spectrum, by
\begin{equation}
\sigma_x^2={1\over(2\pi)^3}\int d^3k\,P(k)W(kx)
\end{equation}

\noindent (see, e.g. Bunn \& White [1997], eqs.~[22] and [23]).
By combining equations~(29), (30), and (31), we get
\begin{equation}
P(k)=V_{\rm box}|\delta_\kbf^{\rm disc}|^2=
{(2\pi)^6\over V_{\rm box}}|\delta_\kbf^{\rm cont}|^2\,.
\end{equation}

\noindent
Both $P(k)$ and $\delta_\kbf^{\rm cont}$ have dimensions of a volume
while $\delta_\kbf^{\rm disc}$ is dimensionless. 
Notice that the form of these 
expressions depends upon the actual definition
of the Fourier Transform, which tends to vary among authors.
%Equation~(23) is taken from Peebles (1980, eq.~[26.1]), but other authors 
%include a factor of $(2\pi/L_{\rm box})^{3/2}=k_0^{3/2}$ on the right hand
%side of this expression. This affects the relationships between 
%$\delta_\kbf^{\rm disc}$, $\delta_\kbf^{\rm cont}$, and the power spectrum.
%Hence, the statement, often heard, that ``the power spectrum is the square
%of the amplitude'' should be instead ``the power spectrum is proportional
%to the square of the amplitude,'' since the proportionality constant 
%depends upon the convention used for the Fourier Transform, and whether
%we are talking about the discrete amplitude or the continuous one.

To set up initial conditions, we lay down the particles on a cubic
lattice, and displace each particle by an amount $\Delta\rbf$ given by
\begin{equation}
\Delta{\rbf}=-i\sum_\kbf
{G_\kbf\delta_\kbf^{\rm disc}\kbf\over k^2}e^{-i\kbf\cdot\rbf}\,,
\end{equation}

\noindent where $\rbf$ is the unperturbed position, 
$\delta_\kbf^{\rm disc}=|\delta_\kbf^{\rm disc}|e^{-i\phi_\kbf}$
is a complex number with amplitude 
$|\delta_\kbf^{\rm disc}|=[P(k)/V_{\rm box}]^{1/2}$
and phase $\phi_\kbf$ chosen randomly
between 0 and $2\pi$ with uniform probability,
and the sum extends over all modes included in the initial conditions 
(see \S2.4). As Efstathiou et al.
(1985) point out, assuming random phases would be sufficient to ensure that
the initial conditions are Gaussian, in the continuous limit (that is,
in an infinite universe). However, this assumption is insufficient in
the discrete limit (that is, in a finite universe with 
periodic boundary conditions). To ensure the Gaussianity of the initial 
conditions, it is necessary, and sufficient, to include the Gaussian factor
$G_\kbf$, a random number chosen from a Gaussian distribution with mean 0
and dispersion 1. This guarantees the initial conditions
are Gaussian, even though there might be a lack of
resolution at some scales; $G_\kbf$ does not change the spectral amplitude
of the fluctuations.

To compute the initial peculiar velocity field, we assume that the initial
time of the calculation is early enough for the perturbation to be
in the linear regime, but late enough so that the linear decaying mode can
be neglected. The initial peculiar velocity of the particles are then 
related to their displacements by
\begin{equation}
{\bf v}_i=\biggl({1\over\delta_+}
{d\delta_+\over dt}\biggr)_{z_i}\Delta\rbf\,,
\end{equation}

\noindent where $z_i$ is the initial redshift of
the simulations, $\Delta\rbf$ is computed using equation~(33),
and $\delta_+$ is the linear growing mode of the perturbation, which
depends upon the cosmological model.
For the Einstein-de~Sitter model ($\Omega_0=1$, $\lambda_0=0$), 
the growing mode is
\begin{equation}
\delta_+(z)=(1+z)^{-1}\,.
\end{equation}

\noindent
For open models ($\Omega_0<1$, $\lambda_0=0$), the growing mode is
\begin{equation}
\delta_+(z)=1+{3\over x}+3\biggl({1+x\over x^3}\biggr)^{1/2}
\ln\big[(1+x)^{1/2}-x^{1/2}\big]
\end{equation}

\noindent (Peebles 1980), where 
\begin{equation}
x=(\Omega_0^{-1}-1)(1+z)^{-1}\,.
\end{equation}

\noindent Finally, for flat models with a cosmological constant
($\Omega_0+\lambda_0=1$), the growing mode is given by
\begin{equation}
\delta_+(z)=\biggl({1\over y}+1\biggr)^{1/2}
\int_0^y{dw\over w^{1/6}(1+w)^{3/2}}
\end{equation}

\noindent (Martel 1991b), where
\begin{equation}
y={\lambda_0\over\Omega_0}(1+z)^{-3}\,.
\end{equation}

\subsection{The Simulations}

We set 
the comoving length of the computational volume $L_{\rm box}$ equal
to $128\,{\rm Mpc}$ (present length units).
The total mass of the system is
$M_{\rm sys}=3H_0^2\Omega_0L_{\rm box}^3/8\pi G=5.821\times10^{17}
\Omega_0h^2{\rm M}_\odot$. We use $N=64^3=262,144$ particles of
mass $M_{\rm part}=M_{\rm sys}/N=2.220\times10^{12}\Omega_0h^2{\rm M}_\odot$.
We solve Poisson's equation on a $128^3$ grid. 
In all simulations, $\epsilon$ and $r_e$ were set equal to
0.3 and 2.7 mesh spacings, respectively. This corresponds, in
physical units, to a comoving softening length $\epsilon=300\,\rm kpc$. 
This is a reasonable value for gravity-only cosmological simulations.
At smaller scales, hydrodynamical effects become important and cannot
be ignored. The dynamical range in length of the algorithm is
$L_{\rm box}/\epsilon=467$.

The ratio of the nyquist wavenumber $k_{\rm nyq}$
to the fundamental wavenumber $k_0$ is $N^{1/3}/2=32$. Hence each component
$k_i$, $i=x$, $y$, $z$,
of the wavenumber can take 65 values; $k_i=n_ik_0$, with
$-32\leq n_i\leq32$. The initial conditions can therefore
represent $65^3=274\,625$ modes. However, the
{\it reality condition} requires that the amplitudes of
modes with equal and opposite wavenumbers are related by
$\delta_{\kbf}^{\rm disc}=(\delta_{-\kbf}^{\rm disc})^*$
[in order for $\delta(\rbf)$ to be real]. Furthermore,
we exclude modes with $|\kbf|=(k_x^2+k_y^2+k_z^2)^{1/2}>k_{\rm nyq}$.
This reduces the actual number of modes 
represented in the initial conditions to $68\,532$.

All simulations start at an initial redshift $z_i=24$. The algorithm produces
``dumps'' (snapshots of the system) at numerous intermediate redshifts, up to
the present. These redshifts where chosen by imposing that the dumps
are equally space in conformal time $\eta$, defined by
$d\eta\equiv a_0dt/a(t)$. We set the difference $\Delta\eta$
between consecutive dumps equal to $L_{\rm box}/c$. Thus, if $t$ and $t'$
are the times corresponding to 2 consecutive dumps, they are related by
\begin{equation}
{L_{\rm box}\over c}=\int_{t'}^t[1+z(t)]dt\,.
\end{equation}

\noindent This particular choice results in most dumps being concentrated near
the present. Typically, about half of the dumps are between redshifts $z=1$ 
and $z=0$. Since the relationship between time and redshift, $z(t)$, is 
model-dependent, the redshifts where dumps are made depend upon
the cosmological parameters $\Omega_0$, $\lambda_0$, and $H_0$
(but not $\sigma_8$). Every simulation also produces a dump at
$z=z_i=24$, and one at $z=0$. The number of dumps per
simulation varies between 44 and 128.

\section{THE DATABASE}

\subsection{The Cosmological Models}

The power spectrum described in
\S2.2 is characterized by 6 independent parameters:
(1)~the density parameter $\Omega_0$, (2)~the contribution~$\Omega_{\rm B0}$ of
the baryonic matter to the density parameter, (3)~the cosmological constant
$\lambda_0$, (4)~the Hubble constant~$H_0$, (5)~the temperature $T_{\rm CMB}$ 
of the Cosmic Microwave Background, and (6)~the tilt $n$ of the power spectrum.
In order to keep the size of the parameter space at a manageable level,
we set $T_{\rm CMB}=2.7\,K$ and $\Omega_{\rm B0}=0.015h^{-2}$, thus 
reducing the dimensionality of the parameter space to 4. Also, the 
normalization of the power spectrum is often described in terms of the
rms density fluctuation $\sigma_8$ at a scale of $8h^{-1}\rm Mpc$. The value
of $\sigma_8$ is a function of the 6 aforementioned parameters. We 
invert this relation, treating $\sigma_8$ as an independent parameter, and
the tilt $n$ as a dependent one. The independent parameters in the database
are therefore $\Omega_0$, $\lambda_0$, $H_0$, and $\sigma_8$.
For each model, we performed up to
3 different simulations, with different realizations of the initial
conditions (this amounts to choosing a different set of random numbers
for the phases $\phi_\kbf$ of the complex numbers $\delta_\kbf^{\rm disc}$,
and the Gaussian factors $G_\kbf$).

An important question was to decide which models should be included in
the database. Our goal here is not to find the ``ultimate model,'' which
provides the best match to current observations. This would defeat the
purpose of having a database, and furthermore, as former supporters of
the Standard Model can appreciate, the ``best model'' can eventually be
proven incorrect by new observations. Our intention is to provide an adequate
coverage of the parameter phase-space. However, we do not want to invest
much effort into simulating models that are considered ``unlikely,'' because
some of the parameters have extreme values. With this in mind, we performed
160 simulations, which provide a broad coverage of the parameter phase-space,
but we favored ``likely'' regions of the parameter phase-space over
``unlikely'' 
ones, by performing more simulations in these regions. For instance,
we consider models with Hubble constant varying in the range
$H_0=50-85\,\rm km\,s^{-1}Mpc^{-1}$, but 139 of the calculations 
(87\%) have a Hubble constant in the more plausible range 
$H_0=65-75\,\rm km\,s^{-1}Mpc^{-1}$. 

The value of the parameters are
given in Table~1 for the entire database
(with $H_0$ in units of $\rm km\,s^{-1}Mpc^{-1}$). 
The first 4 columns contain the
values of the 4 independent parameters $\Omega_0$, $\lambda_0$, $H_0$, 
and $\sigma_8$. The dependent parameter $n$ is in the fifth column.
The sixth and seventh columns contain the number of dumps {\it per simulation}
and the codes of the simulations respectively (see \S3.2). 
The parameter phase-space coverage of
the database is illustrated in Figure~1. The top left panel shows a
projection of the 4-dimensional parameter phase-space onto the 
$\Omega_0-\lambda_0$ plane. The dots indicate the cases for which there are
simulations in the database. The number next to each dot indicates the
number of simulations for that particular combination of $\Omega_0$ and
$\lambda_0$. This panel includes all simulations in the database. 
The top right panel
shows the same projection, but for a subset of the simulations, all simulations
with $H_0=65\rm\,km\,s^{-1}Mpc^{-1}$. The remaining 4 panels show different
projections and different subsets. As we see, the coverage of the 
parameter phase space is quite dense. The biggest ``hole'' is seen in
the $\Omega_0-\lambda_0$ projection (top panels). There are currently no
simulations for open models with a nonzero cosmological constant
($\lambda_0\neq0$ and $\Omega_0+\lambda_0<1$) in the database. As we pointed 
out in \S1.3, these models are certainly worth considering, and we intend
to include such models in the database in the near future. 

Several interesting quantities can be computed directly from the
parameters. One of them is the age of the universe. For $\lambda_0\neq0$
models, $t_0$ is given by
\begin{equation}
t_0={1\over H_0}\int_0^1\biggl[{x\over\lambda_0 x^3+(1-\Omega_0-\lambda_0)x
+\Omega_0x}\biggr]^{1/2}dx
\end{equation}

\noindent
(see, e.g., Martel 1990). It is of course independent of $\sigma_8$. 
Table~2 gives the ages is Gigayears for the various models included
in the database.

Another interesting quantity is $\sigma_8^{\rm clus}$, the value of $\sigma_8$
inferreded from observations of clusters of galaxies.
Using the X-ray temperature distribution function of clusters,
Viana \& Liddle (1996) have produced an empirical formula for
$\sigma_8^{\rm clus}$,
\begin{equation}
\sigma_8^{\rm clus}=0.6\Omega_0^{-C(\Omega_0)}\,,
\end{equation}

\noindent where
\begin{equation}
C(\Omega_0)=\cases{
0.36+0.31\Omega_0-0.28\Omega_0^2\,,&$\lambda_0=0$\,;\cr
0.59-0.16\Omega_0+0.06\Omega_0^2\,,&$\lambda_0=1-\Omega_0$\,.\cr}
\end{equation}

\noindent
Table~3 gives the values of $\sigma_8^{\rm clus}$ 
for the various models included in the database.
Notice that these values do not always match the actual values of $\sigma_8$
used for the calculations (fourth column of Table 1).

\subsection{Nomenclature}

Each cosmological model, that is, each combination of the four parameters
$\Omega_0$, $\lambda_0$, $H_0$, and $\sigma_8$, is identified by a 
two-character code composed of an uppercase letter and a lowercase letter.
For instance, the Einstein-de~Sitter model with 
$H_0=65\,\rm km\,s^{-1}Mpc^{-1}$ and $\sigma_8=1.0$ is identified by
the code {\tt Xa}. The letters were chosen for practical
reasons, and the reader should not try to find some logic in these choices.
Each simulation is identified by a 3-character code, composed of the
two-character code for the model, plus a digit to identified the
simulation. For instance, the three simulations for the {\tt Xa} model are
identified by the codes {\tt Xa1}, {\tt Xa2}, and {\tt Xa3}. The codes for
the entire database are given in the last column of Table~1. Each simulation
produces many output files, or dumps, which are snapshots of the system at 
various redshift. Each file is identified by a 7-character code, which 
consists of the 3-character code of the simulation, an underscore, and
a 3-digit number which identifies the file. For instance, the first
output file created by the simulation {\tt Xa1} is called {\tt Xa1\_001},
and contains a snapshot of the system at the initial redshift $z_i=24$.
The next file created by that simulation is called {\tt Xa1\_002}, and
contains a snapshot at $z=22.079$, and so on. The last file is called 
{\tt Xa1\_058}, and contains a snapshot at the present ($z=0$). The
lists of redshifts where dumps are available can be obtained from the
authors. There is one such list for every combination of 
$\Omega_0$, $\lambda_0$, and $H_0$ included in the database.

\subsection{Conversion to Physical Units}

The positions and velocities stored in the dumps are expressed in
supercomoving variables. They can be converted to physical units
using equations~(5)--(11). The positions $\bf r$ and velocities 
${\bf u}=H(z){\bf r}+{\bf v}$ in physical units are given by
\begin{eqnarray}
{\bf r}&=&{L_{\rm box}\tilde{\bf r}\over1+z}\,,\\
{\bf u}&=&L_{\rm box}\biggl[{H(z)\tilde{\bf r}\over1+z}
+{\Omega_0^{1/2}H_0(1+z)\tilde{\bf v}\over2a_0^{1/2}}\biggr]\,.
\end{eqnarray}

\noindent In these expressions, we have reintroduced the tilde notation for
the supercomoving variables. The expression of ${\bf r}$ is the same for
all models, but the one for ${\bf v}$ is model dependent.
After eliminating $H(z)$ using equation~(12)
and $a_0$ using equations~(13)--(15), we obtain the following expressions:

\noindent (a) Einstein-de~Sitter model ($\Omega_0=1$, $\lambda_0=0$)
\begin{equation}
{\bf u}=H_0L_{\rm box}\biggl[(1+z)^{1/2}\tilde{\bf r}
+{(1+z)\tilde{\bf v}\over2}\biggr]\,.
\end{equation}

\noindent (b) Open models ($\Omega_0<1$, $\lambda_0=0$)
\begin{equation}
{\bf u}=H_0L_{\rm box}\biggl[(1+\Omega_0z)^{1/2}\tilde{\bf r}
+{\Omega_0(1+z)\tilde{\bf v}\over2(1-\Omega_0)^{1/2}}\biggr]\,.
\end{equation}

\noindent (c) Flat models with nonzero cosmological constant
($\Omega_0+\lambda_0=1$)
\begin{equation}
{\bf u}=H_0L_{\rm box}\biggl\{\bigl[\Omega_0(1+z)^3+\lambda_0\bigr]^{1/2}
{\tilde{\bf r}\over1+z}
+{\Omega_0^{2/3}(1+z)\tilde{\bf v}\over2\lambda_0^{1/6}}\biggr\}\,.
\end{equation}

\subsection{Technical Considerations}

The database contains 160 simulations for 68 cosmological models. For
each simulation, there is a dump at $z=z_i=24$ and one at $z=0$, plus
numerous dumps at intermediate redshifts. There is a total of $11\,973$ dumps
in the database. Each dump contains $6N=1\,572\,864$ numbers, the
coordinates of the position and velocity for each particle in the
simulation. These numbers are stored in single precision
(32 bits), although
the simulations themselves were performed in double precision. Each dump
is a binary file (IEEE 754 standard) of size $6.3\rm\,Mb$,
which contains first the $x$-coordinates of all particles, followed by
the $y$-coordinates, the $z$-coordinates, the $v_x$-coordinates, 
the $v_y$-coordinates, and finally the $v_z$-coordinates. Figure~2
shows a sample FORTRAN program that reads a file from the database.
The size of
the entire database is $75.3\rm\,Gb$. The database currently resides on
archival tapes at the High Performance Computing Facility, University
of Texas, where the simulations were performed.

Because of the size of the database, it would be impractical
(if not impossible) to install it on a web site or an anonymous ftp site
where it could be easily retrieved by the user. This might change in the
future, but currently the only way to access the database is to contact the
authors, preferably by E-mail, and send a list of the dumps requested.
Then, the authors and the user can choose the best strategy for transferring
file, according to the computer resources and needs of the user. Requests
should be sent to {\tt database@galileo.as.utexas.edu}.

\section{ANALYSIS OF THE SIMULATIONS}

The goal of this paper is to present the database, and describe its
content. We supplement this description by analyzing
the final state of each simulation, which corresponds to the present. 
We focus on four particular
aspects of the present large-scale structure: the rms density fluctuation,
the two-point correlation function, the moments of the peculiar velocity field,
and the properties of clusters. In this abbreviated version of the
paper, we only present
the analysis of the rms density fluctuation and the two-point
correlation function. The full version of
the paper can be obtained by contacting the authors.

\subsection{RMS Density Fluctuation}

The present rms density fluctuation $\sigma_8$ at scale $8h^{-1}\rm Mpc$ is
treated as an independent parameter in our simulations. 
However, we do not set up the state of the system at present. Instead, we
set up initial conditions at high redshift ($z_i=24$), and evolve the
system numerically all the way to the present. We adjust the initial
conditions in such a way that the density fluctuation at presents ends up
being equal to the desired value of $\sigma_8$. To achieve this, we assume 
that the power spectrum evolves with time according to linear 
perturbation theory
(hence the presence of the factor ${\cal L}^{-2}$ in equation~[16]). Actually,
we do not expect the actual value of $\sigma_8$ to be precisely equal
to the desired one, for several reasons. Let us designate by 
$\sigma_8^{\rm cont}$ the desired value of $\sigma_8$  for each simulation,
that is, the quantity appearing in the fourth column of Table~1. 
The superscript ``cont'' indicates that this the value in the real universe,
where the wavenumber $\kbf$ varies continuously. We are representing the
initial conditions using a finite number of discrete modes, which is
clearly an approximation. We designate by $\sigma_8^{\rm disc}$ the value
of $\sigma_8$ resulting from this approximation. This value is given by
\begin{equation}
(\sigma_8^{\rm disc})^2={\cal L}^2(z_i,0)
\sum_\kbf|\delta_{\kbf,i}^{\rm disc}|^2W(k\ell)\,,
\end{equation}

\noindent where $\ell\equiv8h^{-1}\rm Mpc$.
Hence, $\sigma_8^{\rm disc}$ is computed by summing over all modes
present in the initial conditions, {\it ignoring the Gaussian factor}, and then
extrapolating to the present using linear perturbation theory. 

The introduction of the Gaussian factor in equation~(33) further modifies the
value of $\sigma_8$. We designate by $\sigma_8^{\rm gauss}$ the value
of $\sigma_8$ resulting from the presence of this factor,
\begin{equation}
(\sigma_8^{\rm gauss})^2={\cal L}^2(z_i,0)
\sum_\kbf G_\kbf^2|\delta_{\kbf,i}^{\rm disc}|^2W(k\ell)\,.
\end{equation}

Finally, we designate by $\sigma_8^{\rm num}$ the ``numerical'' value of 
$\sigma_8$, which is the actual rms density fluctuation inside the 
computational volume 
at present, obtained from the numerical simulation. This value
should differ from $\sigma_8^{\rm gauss}$ from several reasons. First,
the numerical algorithm has a finite accuracy, owing to the fact that the time
step is finite and that the gravitational force is softened at short
distances. Second, the evolution of a single mode would never
follow precisely the exact solution when the system is represented by
a finite number of particles. Third, and more importantly, equations~(49)
and~(50) use linear perturbation theory to extrapolate from the initial
conditions to the present, but this can only
be approximate, as mode coupling introduces nonlinear effects at small scale.

We investigated the importance of these various effects by computing
the various values of $\sigma_8$ for all simulations. The value of 
$\sigma_8^{\rm cont}$ is imposed. The values of
$\sigma_8^{\rm disc}$ and $\sigma_8^{\rm gauss}$ can be computed directly
using equations~(49) and~(50) (these values are provided automatically by the
code that generates initial conditions). To evaluate $\sigma_8^{\rm num}$,
we used a direct, somewhat brute-force approach. For each simulation, we
located one million (!) spheres
of radius $8h^{-1}\rm Mpc$ at random locations
inside the computational volume at present, and computed the density
contrast $\delta_{\rm sph}$ inside each sphere, using
\begin{equation}
\delta_{\rm sph}={N_{\rm sph}-\bar N_{\rm sph}\over
\strut\bar N_{\rm sph}}\,,
\end{equation}

\noindent where $N_{\rm sph}$ is the number of particles inside the sphere,
and $\bar N_{\rm sph}$ is the ``mean number,'' given by
\begin{equation}
\bar N_{\rm sph}={4\pi\ell^3N\over3V_{\rm box}}
\end{equation}

\noindent (notice that $\bar N_{\rm sph}$ is not an integer).
The value of $\sigma_8^{\rm num}$ is then given by
\begin{equation}
(\sigma_8^{\rm num})^2=10^{-6}\sum_{\rm all\ spheres}
(\delta_{\rm sph})^2\,.
\end{equation}

We plot these various values of $\sigma_8$ against each others in Figure~3.
The top-left panel shows the effect of discreteness. All dots are located
below the dashed line, indicating that 
$\sigma_8^{\rm disc}<\sigma_8^{\rm cont}$. This is caused not as much by
the discreteness itself as by the fact that modes outside the
range $k_0\leq k\leq k_{\rm nyq}$ are missing in the initial conditions.
Still, the effect is very small, 7\% in the worst case (which happens to
be model {\tt Uc}). As we see on the top right panel, the 
effect of introducing of the
Gaussian factor $G_\kbf$ is quite important, and causes a spread of
order 10\% in the value of $\sigma_8$. This is primarily an effect 
of undersampling. The modes with wavenumbers comparable to $k_0$ are very few,
but contribute significantly to $\sigma_8$. This is consequence of the fact
that our computational volume is actually too small to constitute a
``fair sample'' of the universe. With a larger volume, there would
still be very few modes with wavenumbers of order $k_0$, but these modes
would be farther from the peak of the power spectrum, and therefore would contribute less to $\sigma_8$. The bottom left panel shows the effect of
actually performing the simulation. The values of $\sigma_8^{\rm gauss}$
and $\sigma_8^{\rm num}$ are comparable for small $\sigma_8$, but differ
at large $\sigma_8$ where nonlinear effects become important. This panel shows
that (1) the onset of nonlinearity occurs at $\sigma_8\sim0.6$, (2)
The effect of nonlinearity on the value of $\sigma_8$ is small, of
order 10\%, and (3) {\it the effect can go either way}, it can either
increase or decrease the value of $\sigma_8$, and the occurrences of these
two cases are comparable. 

The bottom right panel shows the combined
effect of discreteness, Gaussian factor, and nonlinearity. The
spread is quite large. The value $\sigma_8^{\rm num}$ that comes out of the
simulation can differ by as much as 20\% from the value $\sigma_8^{\rm cont}$
we {\it intended} to obtain, that is, the values given in the fourth column of
Table~1. The reader should be aware of this fact when selecting a particular
model from the database. Consequently, we listed in
Table~4 the values of $\sigma_8^{\rm num}$ for all simulations.

\subsection{2-point Correlation Function}

The distribution of galaxies in the universe can be described statistically
using N-point correlation functions. The first and most important of
these functions is the 2-point correlation function $\xi(r)$, which measures 
the excess probability of finding two galaxies separated by a distance $r$.
The 2-point correlation function can be estimated from the particle 
distributions by locating spherical shells around particles and counting the
number of particles inside these shells. If $N(r_1,r_2)$ is the number
of particles inside a spherical shell of inner radius $r_1$ and outer radius
$r_2$ centered on a given particles, and $\langle N(r_1,r_2)\rangle$
is the average of $N(r_1,r_2)$ over all particles, then by definition
\begin{equation}
\langle N(r_1,r_2)\rangle={4\pi(r_2^3-r_1^3)n\over3}+4\pi n\int_{r_1}^{r_2}
\xi(r)r^2dr\,.
\end{equation}

\noindent where $n$ is the number density of particles. The first term in
equation~(54) gives the correct answer in the case of a uniform distribution.
The second term represents the effect of the correlation. We can solve this
equation for $\xi$ (see Martel 1991a). After some algebra, we get
\begin{equation}
\xi(x)={1\over r^3(x)\ln10}{d\over dx}
\biggl[{\langle N(r(x),r_1)\rangle\over4\pi n}-{r^3(x)-r_1^3\over3}\biggr]\,,
\end{equation}

\noindent where $x\equiv\log_{10}r$. To compute $\xi$, we 
first evaluate $\langle N(r(x),r_1)\rangle$ by computing the spacing between
all pairs of particles, and counting pairs in bins equally spaced in $x$. We
then compute the derivative in equation~(55) using a standard five-point finite
difference operator. We computed $\xi(r)$ at present for all simulations
in the database. The results are plotted in Figures~4--7. For models with
more than one simulation (most have three), we averaged the curves. They
were actually so similar, in all cases, that error bars in Figures~4--7
would be too small to be seen. This shows that the 2-point correlation function
depends essentially upon the cosmological model, with very little dependence
upon the particular realization of the initial conditions. Each panel
in Figures~4--7 corresponds to a particular combination of 
$\Omega_0$, $\lambda_0$, and $H_0$, with different values of $\sigma_8$ 
represented by different curves. The dashed lines show the observed galaxy 
2-point correlation function,
\begin{equation}
\xi(r)=\biggl({r\over5.4h^{-1}{\rm Mpc}}\biggr)^{-1.77}
\end{equation}

\noindent (Peebles 1993, eq.~[7.32]). Models with $\sigma_8=0.3$ fail
to reproduce the observed correlation function on three counts. The slope of
$\xi(r)$ is too shallow, the amplitude is too small,
and there tend to be a ``kink'' in the
correlation function at separations of order $1-3\,\rm Mpc$. As 
$\sigma_8$ increases, the kink goes away, and the amplitude and slope increase.
Models with
$\sigma_8=0.8$ provide the best fit to the observations, almost
independently of the values of the other parameters! For larger values of
$\sigma_8$, the slope and amplitude are too large. 

All curves have a shoulder at small separations, where the slope drops
significantly. Martel (1991a) argues that this is a consequence of the 
softening of the force at small distances. The effect of this softening
is to ``take'' pairs of particles that would have a separation $r$
less than the softening length $\epsilon=300\,\rm kpc$ in the absence of
softening, and transfer them to separations $r\gtrsim\epsilon$, resulting in
a flattening of the correlation function. To illustrate this,
we indicated in all panels of Figures~4--7 the location of the softening
length $\epsilon$ by a thick line. This line is located right in the middle
of the ``shoulder'' for all curves. With higher force resolution, pairs
that are now located at separations $r\gtrsim\epsilon$ would be located instead
at separations $r\lesssim\epsilon$, and the fit to the observed slope would 
be improved.

\section{SUMMARY AND PROSPECTS}

Using a P$^3$M algorithm with $64^3$ particles, we have performed 160
cosmological simulations, for 68 cosmological models. This constitutes
the largest database of cosmological simulations ever assembled. 
We covered a four-dimensional parameter phase space by varying the
density parameter $\Omega_0$, the cosmological constant $\lambda_0$,
the Hubble constant $H_0$, and the rms density fluctuation $\sigma_8$.
We are making this database available to the astronomical community. We also
performed a limited analysis of the simulations. Our results are
the following:

\noindent (1) The present rms density fluctuation $\sigma_8^{\rm num}$
differs from the one expected by linearly extrapolating the initial
power spectrum to the present, because of the combined effects of having
a finite number of modes in the initial conditions, introducing a
Gaussian factor in the initial conditions, and having nonlinear coupling
between modes. The first effect is negligible. The second effect is of
order 10\% or less, but this probably depends upon the size of the
computational volume [the one used for the simulations, $(128\,\rm Mpc)^3$,
is a little too small to constitute a fair sample of the universe]. The effect
of nonlinearity is negligible for $\sigma_8<0.6$, and of order 10\% or
less for larger $\sigma_8$. This can go either way: nonlinearities can either
increase of decrease $\sigma_8$ relative to what linear theory predicts.

\noindent (2) The observed two-point correlation function $\xi(r)$ is
well reproduced by models with $\sigma_8=0.8$, {\it nearly independently
of the values of the parameters $\Omega_0$, $\lambda_0$, and $H_0$}.
For models with $\sigma_8<0.8$, the correlation function is too small,
its slope is too shallow, and it has a kink at separations 
$r=1-3\,\rm Mpc$. For models with $\sigma_8>0.8$, the correlation function
is too large and its slope is too steep.

\noindent (3) At small separations, $r<1\,\rm Mpc$, 
the velocity moments satisfy the relations $|V_{\rm R}|\approx H_0r$
and $V_{\rm PP}\approx2^{1/2}V_{\rm PL}$, indicating that small
clusters have reached virial equilibrium. At larger separations, $|V_{\rm R}|$
increases above the Hubble velocity, indicating that clusters are accreting
matter from the field. The velocity moments depend essentially
upon $\Omega_0$ and $\sigma_8$, and not $\lambda_0$ and $H_0$.
The pairwise particle velocity dispersions are much 
larger than the observed pairwise galaxy velocity dispersion, except for
models with $\Omega_0=0.2$ and $\sigma_8\leq0.4$. 
But if the velocity dispersion
of galaxies is biased relative to the velocity dispersion of dark matter,
then models with larger values of $\Omega_0$ or $\sigma_8$
can be reconciled with observations.

\noindent (4) The multiplicity functions are decreasing for small values of
for models with $\sigma_8\sim0.3$. At larger values of $\sigma_8$, the 
multiplicity functions have a horizontal plateau, whose length increases 
with $\sigma_8$. For models with $\sigma_8>0.9$, the multiplicity functions
have a $\cup$ shape which results from the merging of intermediate-size 
clusters. For all models, clusters have densities
in the range $100\bar\rho_0-1000\bar\rho_0$. 
A simple analytical model suggest that
clusters have a density $\rho\sim178\bar\rho$ when they reach virial
equilibrium. Our results suggest that many clusters have reached 
that equilibrium in the past, when $\bar\rho$ was larger than $\bar\rho_0$
(this could be checked by performing a cluster analysis on earlier dumps).
The spin parameters $\lambda$ are in the range $0.008-0.2$, with the
median near 0.05, and the distributions of elongations favors
prolate shapes ($e_2>e_1$) over oblate shapes ($e_1<e_2$). 
These results indicate the absence of rotationally 
supported disks in these simulations.

The database is growing. We are currently adding new simulations
to the original 160 simulations described in this paper.
There are at least
seven different motivations for performing additional simulations. 

\noindent
(1) Additional Simulations for the same Models: 
For the sake of providing a good coverage of the
parameter phase-space, we have limited the number of simulations per model
to 3 or less. We can perform additional simulations for models already 
included in the database, if there is a need for doing so. 
This could be the case if,
for some reason, a particular model (that is, a particular combination
of $\Omega_0$, $\lambda_0$, $H_0$, and $\sigma_8$) becomes particularly
interesting, and deserves more scrutiny. Also, having more simulations
per model has the virtue of improving the statistics. For instance, the
size of the error bars in Figures~8--12 would be reduced if we had more
than 3 simulations per model. Finally, for gravitational lensing
simulations, it is necessary to combine dumps generated
by different simulations, and having more simulations can be desirable
(see, e.g. Premadi, Martel, \& Matzner 1998).

\noindent
(2) Different Box Size: All simulations included in the database
were performed using a computational box of size $L_{\rm box}=128\,\rm Mpc$.
As described in \S2.4, the softening length is comparable to the scale
where nongravitational effects become important. Hence, there is no
reason to consider smaller box sizes, unless we want to use {\it fewer}
particles. There are, however, reasons for considering larger boxes. 
As we pointed out in \S4.1, a box of size $128\,\rm Mpc$ is too small to
constitute a ``fair sample'' of the universe. Using boxes of size
$256\,\rm Mpc$ or even $512\,\rm Mpc$ would certainly provide a better,
``fairer'' description of the large-scale structure, even with the same 
number of particles. 

\noindent
(3) Larger Number of Particles:
The is no point increasing the number of particles as long as we
keep the box size at $128\,\rm Mpc$, since the resolution would be increased
at scales where nongravitational effects are important. However, if larger
boxes are used, the number of particles can be increased accordingly in
order to maintain the resolution of the algorithm at small scale. 
If we continue to adopt $300\,\rm kpc$ as the resolution scale of the
algorithm, simulations in $(256\,\rm Mpc)^3$ and $(512\,\rm Mpc)^3$ boxes 
could be performed with $128^3$ and $256^3$ particles, respectively.  

\noindent
(4) New Background Models:
The 4-dimensional parameter phase-space considered in this
paper is quite large, and the set of 68 
cosmological models
included in the database covers a small fraction of it. There
are several ``holes'' in the projections shown in Figure~1. 
In particular, there are no simulations for open models
with a nonzero cosmological constant ($\lambda_0\neq0$,
$\Omega_0+\lambda_0<1$).
Simulations for additional background models 
could be added to the database, either to provide
a better coverage of the parameter phase-space, or because there is
a particular model we are interested in, ``we'' designating either
the authors, or other researchers sending us a special request.
Actually, the original database contained only 151 simulations for
65 cosmological models. Following a special request by Hamana (1998),
we added 9 simulations to the database, for 3 new models: 
{\tt Ea}, {\tt Pa}, and {\tt Xg}. 

\noindent
(5) Additional Parameters:
The current database covers a 4-parameter phase space, because we held
the CMB temperature $T_{\rm CMB}$ and the baryon density parameter
$\Omega_{\rm B0}$ at values of 2.7 and $0.015h^{-2}$, 
respectively. The CMB temperature is known so accurately that treating
it as a variable parameter
would be pointless. This is not the case for the baryon
density parameter. According to primordial nucleosynthesis,
the quantity $\Omega_{\rm B0}h^2$ has an allowed range 
from 0.01 to 0.026 (Krauss \& Kernan 1995; Copi et al. 1995; Krauss 1998).
Furthermore, X-ray observations of clusters of galaxies suggest
that the ratio of gas mass to dark matter mass in these clusters exceeds the
mean value in the universe, a phenomenon known as ``the baryon catastrophe''
(Briel et al. 1992; White et al. 1993; Martel et al. 1994).
There is at present no definitive explanation for this phenomenon, but
one possible explanation is that primordial nucleosynthesis is 
somehow incorrect,
and predicts a value of $\Omega_{\rm B0}$ which is too small. 

\noindent
(6) Different Components:
All simulations in the database used a CDM power spectrum as initial 
conditions. There are, however, several other models that 
constitute interesting alternatives to the CDM model, which
could be added to the database. One of them is the
Hot Dark Matter model (HDM), though this model has fallen out of favor
in recent years, due to its inability to form galaxies inside deep voids
such as Bo\"otes. A more interesting alternative is the mixed Cold + Hot
Dark Matter model (CHDM), which contains both a cold dark matter component
and a massive neutrino component. This model introduces one additional
parameter, the contribution $\Omega_{\nu0}$ of the neutrinos to the mean energy
density of the universe.  

\noindent
(7) Different Cosmologies:
The cosmological models included in the database contain only 
non-relativistic matter and a nonzero cosmological
constant. It would be very interesting to consider models with other
components. Possible candidates include domain walls, cosmic
strings, or relativistic particles (Fry 1985; Charlton \& Turner 1987;
Silveira \& Waga 1994; Martel 1995; Martel \& Shapiro 1998). Recently, 
these various candidates have
been combined into a single concept called ``quintessence''
(Caldwell, Dave, \& Steinhardt 1998).
The effects of these various components is twofold: First, the presence
of these components modifies the expansion rate of the universe and the
growth rate of density perturbations, thus changing the history of 
large-scale structure formation. Second, 
they might affect the shape and normalization
of the primordial power spectrum, in ways that remain to be determined
(none of these models were considered by Bunn \& White [1997]).

\acknowledgments
 
This work benefited from stimulating discussions with Paul Shapiro.
We are pleased to acknowledge the support of NASA Grants NAG5-2785,
NAG5-7363, and NAG5-7821,
NSF Grants PHY93~10083, PHY98~00725 and ASC~9504046,
the University of Texas High Performance Computing Facility
through the office of the vice president for research.
HM acknowledges the support of a fellowship provided by the
Texas Institute for Computational and Applied Mathematics.

\appendix

\section{Calculation of $\sigma_x^{\rm disc}$}

The mass inside a sphere centered at $\rbf_0$ is given by
\begin{equation}
M(\rbf_0)=\int_{{\rm sph}(\rbf_0)}\bar\rho_{\rm com}(1+\delta)d^3r
=\bar\rho_{\rm com}\biggl[V_{\rm sph}+
\int_{{\rm sph}(\rbf_0)}d^3r\sum_\kbf
\delta_\kbf^{\rm disc}e^{-i\kbf\cdot\rbf}\biggr]\,,
\end{equation}

\noindent where $\bar\rho_{\rm com}$ is the average comoving density,
$V_{\rm sph}$ is the volume of the sphere, and the integral is computed
over that volume. The relative mass excess in the sphere is given by
\begin{equation}
{\Delta M\over M}(\rbf_0)={1\over V_{\rm sph}}
\int_{{\rm sph}(\rbf_0)}d^3r\sum_\kbf\delta_\kbf^{\rm disc}e^{-i\kbf\cdot\rbf}
\,.
\end{equation}

\noindent We introduce the following change of variables,
\begin{equation}
\rbf=\rbf_0+\ybf\,.
\end{equation}

\noindent In $\ybf$-space, the sphere is now located at the origin, and 
equation~(A3) becomes
\begin{equation}
{\Delta M\over M}(\rbf_0)={1\over V_{\rm sph}}
\int_{{\rm sph}(0)}d^3y\sum_\kbf\delta_\kbf^{\rm disc}e^{-i\kbf\cdot\rbf_0}
e^{-i\kbf\cdot\ybf}\,.
\end{equation}

\noindent We now square this expression, and get
\begin{equation}
\biggl({\Delta M\over M}\biggr)^2(\rbf_0)={9\over16\pi^2x^6}
\Biggl[\int_{\rm sph(0)}d^3y\sum_\kbf\delta_\kbf^{\rm disc}
e^{-i\kbf\cdot\rbf_0}e^{-i\kbf\cdot\ybf}\Biggr]
\Biggl[\int_{\rm sph(0)}d^3z\sum_{\kbf'}\delta_{\kbf'}^{\rm disc}
e^{-i\kbf'\cdot\rbf_0}e^{-i\kbf'\cdot\zbf}\Biggr]\,,
\end{equation}

\noindent where $x$ is the radius of the sphere. The rms density
contrast at scale $x$ is obtained by averaging the above expression over 
all possible locations of the sphere inside the computational box,
\begin{eqnarray}
\sigma_x^2&\equiv&
\Biggl\langle\biggl({\Delta M\over M}\biggr)^2\Biggr\rangle_{V_{\rm box}}
={1\over V_{\rm box}}\int_{V_{\rm box}}d^3r_0
\biggl({\Delta M\over M}\biggr)^2(\rbf_0)\nonumber \\
&=&{1\over V_{\rm box}}{9\over16\pi^2x^6}\int_{V_{\rm box}}d^3r_0
\int_{\rm sph(0)}d^3y\int_{\rm sph(0)}d^3z\sum_\kbf\sum_{\kbf'}
\delta_\kbf^{\rm disc}\delta_{\kbf'}^{\rm disc}
e^{-i\kbf\cdot\ybf}e^{-i\kbf'\cdot\zbf}
e^{-i(\kbf+\kbf')\cdot\rbf_0}\,.
\end{eqnarray}

\noindent The integral over $V_{\rm box}$ reduces to
\begin{equation}
\int_{V_{\rm box}}d^3r_0e^{-i(\kbf+\kbf')\cdot\rbf_0}=
V_{\rm box}\delta_{\kbf,-\kbf'}\,.
\end{equation}

\noindent We substitute this expression in equation~(A6), and use the
Kronecker~$\delta$ to eliminate the summation over $\kbf'$. Equation~(A6)
reduces to
\begin{equation}
\sigma_x^2={9\over16\pi^2x^6}\sum_\kbf|\delta_\kbf^{\rm disc}|^2
\Biggl[\int_{\rm sph(0)}d^3y\,e^{-i\kbf\cdot\ybf}\Biggr]^2\,.
\end{equation}

\noindent The remaining integral can be evaluated easily. Equation~(A8)
reduces to
\begin{equation}
\sigma_x^2=\sum_\kbf|\delta_\kbf^{\rm disc}|^2W(kx)\,,
\end{equation}

\noindent where
\begin{equation}
W(y)\equiv{9\over y^6}(\sin y-y\cos y)^2\,.
\end{equation}

\noindent 
Using equation~(26), we can rewrite this expression in an integral form,
\begin{equation}
\sigma_x^2={V_{\rm box}\over(2\pi)^3}\int d^3k
|\delta_\kbf^{\rm disc}|^2W(kx)\,.
\end{equation}

\clearpage

\begin{deluxetable}{ccccccl}
%\footnotesize
\tablecaption{Parameters for the Entire Database}
\tablewidth{0pt}
\tablehead{
\colhead{$\Omega_0$} & \colhead{$\lambda_0$} & \colhead{${\rm H}_0$} &
\colhead{$\sigma_8$} & \colhead{$n$} & \colhead{\# of dumps} &
\colhead{Codes}
}
\startdata
0.20 & 0.00 & 55 & 0.3 & 1.2187 & 101 & {\tt Gb1, Gb2, Gb3} \nl
0.20 & 0.00 & 60 & 0.3 & 1.1539 &  92 & {\tt Ae1}           \nl
0.20 & 0.00 & 65 & 0.3 & 1.0966 &  85 & {\tt Ac1, Ac2, Ac3} \nl
0.20 & 0.00 & 65 & 0.5 & 1.3188 &  85 & {\tt Aa1, Aa2, Aa3} \nl
0.20 & 0.00 & 70 & 0.3 & 1.0454 &  79 & {\tt Af1}           \nl
0.20 & 0.00 & 75 & 0.3 & 0.9993 &  74 & {\tt Ad1, Ad2, Ad3} \nl
0.20 & 0.00 & 75 & 0.4 & 1.1228 &  74 & {\tt Ta1, Ta2, Ta3} \nl
0.20 & 0.00 & 75 & 0.5 & 1.2190 &  74 & {\tt Ab1, Ab2, Ab3} \nl
0.20 & 0.00 & 75 & 0.6 & 1.2979 &  74 & {\tt Tb1, Tb2, Tb3} \nl
0.20 & 0.00 & 75 & 0.7 & 1.3648 &  74 & {\tt Tc1, Tc2, Tc3} \nl
0.20 & 0.00 & 85 & 0.3 & 0.9191 &  65 & {\tt Hb1, Hb2, Hb3} \nl
\tableline
0.20 & 0.80 & 55 & 0.8 & 1.2057 & 128 & {\tt Gc1, Gc2, Gc3} \nl
0.20 & 0.80 & 60 & 0.6 & 0.9948 & 117 & {\tt Ke1}           \nl
0.20 & 0.80 & 65 & 0.6 & 0.9326 & 108 & {\tt Kc1, Kc2, Kc3} \nl
0.20 & 0.80 & 65 & 0.7 & 1.0062 & 108 & {\tt Ua1, Ua2, Ua3} \nl
0.20 & 0.80 & 65 & 0.8 & 1.0702 & 108 & {\tt Ka1, Ka2, Ka3} \nl
0.20 & 0.80 & 65 & 0.9 & 1.1269 & 108 & {\tt Ub1, Ub2, Ub3} \nl
0.20 & 0.80 & 65 & 1.0 & 1.1568 & 108 & {\tt Uc1, Uc2, Uc3} \nl
0.20 & 0.80 & 70 & 0.6 & 0.8771 & 101 & {\tt Kf1}           \nl
0.20 & 0.80 & 75 & 0.6 & 0.8273 &  94 & {\tt Kd1, Kd2, Kd3} \nl
0.20 & 0.80 & 75 & 0.8 & 0.9629 &  94 & {\tt Kb1, Kb2, Kb3} \nl
0.20 & 0.80 & 85 & 0.8 & 0.8749 &  83 & {\tt Hc1, Hc2, Hc3} \nl
\tableline
0.30 & 0.00 & 75 & 0.85 & 1.1748 & 68 & {\tt Ea1, Ea2, Ea3} \nl
\tableline
0.30 & 0.70 & 75 & 0.9 & 0.8796 &  81 & {\tt Pa1, Pa2, Pa3} \nl
\tableline
0.35 & 0.00 & 60 & 0.6 & 1.0670 &  82 & {\tt De1} \nl
0.35 & 0.00 & 65 & 0.6 & 1.0167 &  76 & {\tt Da1} \nl
0.35 & 0.00 & 65 & 0.8 & 1.1428 &  76 & {\tt Dc1} \nl
0.35 & 0.00 & 70 & 0.6 & 0.9718 &  71 & {\tt Df1} \nl
0.35 & 0.00 & 75 & 0.6 & 0.9314 &  66 & {\tt Db1} \nl
0.35 & 0.00 & 75 & 0.8 & 1.0561 &  66 & {\tt Dd1} \nl
\tableline
0.35 & 0.65 & 60 & 0.7 & 0.8614 &  95 & {\tt Ne1} \nl
0.35 & 0.65 & 65 & 0.7 & 0.8098 &  88 & {\tt Na1} \nl
0.35 & 0.65 & 65 & 0.9 & 0.9251 &  88 & {\tt Nc1} \nl
0.35 & 0.65 & 70 & 0.7 & 0.7640 &  82 & {\tt Nf1} \nl
0.35 & 0.65 & 75 & 0.7 & 0.7228 &  76 & {\tt Nb1} \nl
0.35 & 0.65 & 75 & 0.9 & 0.8363 &  76 & {\tt Nd1} \nl
\tableline
0.50 & 0.00 & 60 & 0.8 & 0.9912 &  76 & {\tt Be1}           \nl 
0.50 & 0.00 & 65 & 0.8 & 0.9457 &  70 & {\tt Bc1, Bc2, Bc3} \nl
0.50 & 0.00 & 65 & 1.0 & 1.0439 &  70 & {\tt Ba1, Ba2, Ba3} \nl
0.50 & 0.00 & 70 & 0.8 & 0.9051 &  65 & {\tt Bf1}           \nl 
0.50 & 0.00 & 75 & 0.8 & 0.8686 &  61 & {\tt Bd1, Bd2, Bd3} \nl
0.50 & 0.00 & 75 & 1.0 & 0.9656 &  61 & {\tt Bb1, Bb2, Bb3} \nl
\tableline
0.50 & 0.50 & 60 & 0.8 & 0.8264 &  83 & {\tt Le1}           \nl 
0.50 & 0.50 & 65 & 0.8 & 0.7808 &  77 & {\tt Lc1, Lc2, Lc3} \nl
0.50 & 0.50 & 65 & 1.0 & 0.8807 &  77 & {\tt La1, La2, La3} \nl
0.50 & 0.50 & 70 & 0.8 & 0.7403 &  71 & {\tt Lf1}           \nl 
0.50 & 0.50 & 75 & 0.8 & 0.7049 &  66 & {\tt Ld1, Ld2, Ld3} \nl
0.50 & 0.50 & 75 & 1.0 & 0.8024 &  66 & {\tt Lb1, Lb2, Lb3} \nl
\tableline
0.70 & 0.00 & 65 & 0.9 & 0.8461 &  64 & {\tt Cc1, Cc2, Cc3} \nl
0.70 & 0.00 & 65 & 1.1 & 0.9346 &  64 & {\tt Ca1, Ca2, Ca3} \nl
0.70 & 0.00 & 75 & 0.9 & 0.7773 &  56 & {\tt Cd1, Cd2, Cd3} \nl
0.70 & 0.00 & 75 & 1.1 & 0.8648 &  56 & {\tt Cb1, Cb2, Cb3} \nl
\tableline
0.70 & 0.30 & 65 & 0.9 & 0.7720 &  67 & {\tt Mc1, Mc2, Mc3} \nl
0.70 & 0.30 & 65 & 1.1 & 0.8601 &  67 & {\tt Ma1, Ma2, Ma3} \nl
0.70 & 0.30 & 75 & 0.9 & 0.7042 &  58 & {\tt Md1, Md2, Md3} \nl
0.70 & 0.30 & 75 & 1.1 & 0.7912 &  58 & {\tt Mb1, Mb2, Mb3} \nl
\tableline
1.00 & 0.00 & 50 & 0.5 & 0.5836 &  75 & {\tt Xg1, Xg2, Xg3} \nl
1.00 & 0.00 & 55 & 1.0 & 0.8465 &  69 & {\tt Ga1, Ga2, Ga3} \nl
1.00 & 0.00 & 60 & 1.0 & 0.8057 &  63 & {\tt Xe1}           \nl
1.00 & 0.00 & 65 & 0.9 & 0.7234 &  58 & {\tt Sa1, Sa2, Sa3} \nl
1.00 & 0.00 & 65 & 1.0 & 0.7698 &  58 & {\tt Xa1, Xa2, Xa3} \nl
1.00 & 0.00 & 65 & 1.1 & 0.8120 &  58 & {\tt Sb1, Sb2, Sb3} \nl
1.00 & 0.00 & 65 & 1.2 & 0.8506 &  58 & {\tt Xc1, Xc2, Xc3} \nl
1.00 & 0.00 & 65 & 1.3 & 0.8861 &  58 & {\tt Sc1, Sc2, Sc3} \nl
1.00 & 0.00 & 70 & 1.0 & 0.7380 &  54 & {\tt Xf1}           \nl
1.00 & 0.00 & 75 & 1.0 & 0.7094 &  50 & {\tt Xb1, Xb2, Xb3} \nl
1.00 & 0.00 & 75 & 1.2 & 0.7893 &  50 & {\tt Xd1, Xd2, Xd3} \nl
1.00 & 0.00 & 85 & 1.0 & 0.6605 &  44 & {\tt Ha1, Ha2, Ha3} \nl
\enddata
\end{deluxetable}
\clearpage

\begin{deluxetable}{cccr}
%\footnotesize
\tablecaption{Age of the Universe in Gigayears}
\tablewidth{0pt}
\tablehead{
\colhead{$\Omega_0$} & \colhead{$\lambda_0$} & \colhead{${\rm H}_0$} 
& \colhead{$t_0$}
}
\startdata
0.20 & 0.00 & 55 & 15.05 \nl
0.20 & 0.00 & 60 & 13.79 \nl
0.20 & 0.00 & 65 & 12.73 \nl
0.20 & 0.00 & 70 & 11.82 \nl
0.20 & 0.00 & 75 & 11.04 \nl
0.20 & 0.00 & 85 &  9.74 \nl
\tableline
0.20 & 0.80 & 55 & 19.13 \nl
0.20 & 0.80 & 60 & 17.53 \nl
0.20 & 0.80 & 65 & 16.19 \nl
0.20 & 0.80 & 70 & 15.03 \nl
0.20 & 0.80 & 75 & 14.03 \nl
0.20 & 0.80 & 85 & 12.38 \nl
\tableline
0.30 & 0.00 & 75 & 10.54 \nl
\tableline
0.30 & 0.70 & 75 & 12.57 \nl
\tableline
0.35 & 0.00 & 60 & 12.92 \nl
0.35 & 0.00 & 65 & 11.93 \nl
0.35 & 0.00 & 70 & 11.08 \nl
0.35 & 0.00 & 75 & 10.34 \nl
\tableline
0.35 & 0.65 & 60 & 15.04 \nl
0.35 & 0.65 & 65 & 13.88 \nl
0.35 & 0.65 & 70 & 12.89 \nl
0.35 & 0.65 & 75 & 12.03 \nl
\tableline
0.50 & 0.00 & 60 & 12.28 \nl 
0.50 & 0.00 & 65 & 11.34 \nl
0.50 & 0.00 & 70 & 10.53 \nl 
0.50 & 0.00 & 75 &  9.82 \nl
\tableline
0.50 & 0.50 & 60 & 13.54 \nl 
0.50 & 0.50 & 65 & 12.50 \nl
0.50 & 0.50 & 70 & 11.61 \nl 
0.50 & 0.50 & 75 & 10.83 \nl
\tableline
0.70 & 0.00 & 65 & 10.72 \nl
0.70 & 0.00 & 75 &  9.29 \nl
0.70 & 0.30 & 65 & 11.26 \nl
0.70 & 0.30 & 75 &  9.76 \nl
\tableline
1.00 & 0.00 & 50 & 13.04 \nl
1.00 & 0.00 & 55 & 11.85 \nl
1.00 & 0.00 & 60 & 10.86 \nl
1.00 & 0.00 & 65 & 10.03 \nl
1.00 & 0.00 & 70 &  9.31 \nl
1.00 & 0.00 & 75 &  8.96 \nl
1.00 & 0.00 & 85 &  7.67 \nl
\enddata
\end{deluxetable}

\clearpage

\begin{deluxetable}{ccc}
%\footnotesize
\tablecaption{RMS Density Fluctuations from X-ray Clusters}
\tablewidth{0pt}
\tablehead{
\colhead{$\Omega_0$} & \colhead{$\lambda_0$} & \colhead{$\sigma_8^{\rm clus}$}
}
\startdata
0.20 & 0.00 & 1.162 \nl
0.30 & 0.00 & 1.004 \nl
0.35 & 0.00 & 0.946 \nl
0.50 & 0.00 & 0.827 \nl
0.70 & 0.00 & 0.702 \nl
1.00 & 0.00 & 0.600 \nl
\tableline
0.20 & 0.08 & 1.479 \nl
0.30 & 0.70 & 1.160 \nl
0.35 & 0.65 & 1.059 \nl
0.50 & 0.50 & 0.863 \nl
0.70 & 0.30 & 0.719 \nl
1.00 & 0.00 & 0.600 \nl
\enddata
\end{deluxetable}

\clearpage

\begin{deluxetable}{cccccccc}
%\footnotesize
\tablecaption{Values of $\sigma_8^{\rm num}$ for all Simulations}
\tablewidth{0pt}
\tablehead{
\colhead{Codes} & \colhead{$\sigma_8^{\rm num}$} &
\colhead{Codes} & \colhead{$\sigma_8^{\rm num}$} &
\colhead{Codes} & \colhead{$\sigma_8^{\rm num}$} &
\colhead{Codes} & \colhead{$\sigma_8^{\rm num}$}
}
\startdata
{\tt Aa1}& 0.522 & {\tt Da1}& 0.631 & {\tt Kf1}& 0.741 & {\tt Sb2}& 1.033 \nl
{\tt Aa2}& 0.575 & {\tt Db1}& 0.712 & {\tt La1}& 0.903 & {\tt Sb3}& 1.101 \nl
{\tt Aa3}& 0.450 & {\tt Dc1}& 0.807 & {\tt La2}& 1.099 & {\tt Sc1}& 1.013 \nl
{\tt Ab1}& 0.579 & {\tt Dd1}& 0.881 & {\tt La3}& 1.258 & {\tt Sc2}& 1.245 \nl
{\tt Ab2}& 0.479 & {\tt De1}& 0.606 & {\tt Lb1}& 0.868 & {\tt Sc3}& 1.332 \nl
{\tt Ab3}& 0.475 & {\tt Df1}& 0.595 & {\tt Lb2}& 0.990 & {\tt Ta1}& 0.379 \nl
{\tt Ac1}& 0.269 & {\tt Ea1}& 0.886 & {\tt Lb3}& 0.966 & {\tt Ta2}& 0.395 \nl
{\tt Ac2}& 0.347 & {\tt Ea2}& 0.902 & {\tt Lc1}& 0.689 & {\tt Ta3}& 0.425 \nl
{\tt Ac3}& 0.280 & {\tt Ea3}& 0.846 & {\tt Lc2}& 0.731 & {\tt Tb1}& 0.600 \nl
{\tt Ad1}& 0.321 & {\tt Ga1}& 1.024 & {\tt Lc3}& 0.689 & {\tt Tb2}& 0.618 \nl
{\tt Ad2}& 0.308 & {\tt Ga2}& 1.094 & {\tt Ld1}& 0.843 & {\tt Tb3}& 0.601 \nl
{\tt Ad3}& 0.289 & {\tt Ga3}& 0.821 & {\tt Ld2}& 0.709 & {\tt Tc1}& 0.773 \nl
{\tt Ae1}& 0.268 & {\tt Gb1}& 0.352 & {\tt Ld3}& 0.841 & {\tt Tc2}& 0.694 \nl
{\tt Af1}& 0.326 & {\tt Gb2}& 0.283 & {\tt Le1}& 0.740 & {\tt Tc3}& 0.681 \nl
{\tt Ba1}& 0.863 & {\tt Gb3}& 0.276 & {\tt Lf1}& 0.782 & {\tt Ua1}& 0.776 \nl
{\tt Ba2}& 1.036 & {\tt Gc1}& 0.809 & {\tt Ma1}& 1.145 & {\tt Ua2}& 0.595 \nl
{\tt Ba3}& 0.994 & {\tt Gc2}& 0.749 & {\tt Ma2}& 1.121 & {\tt Ua3}& 0.643 \nl
{\tt Bb1}& 0.828 & {\tt Gc3}& 0.851 & {\tt Ma3}& 1.007 & {\tt Ub1}& 0.707 \nl
{\tt Bb2}& 1.030 & {\tt Ha1}& 0.940 & {\tt Mb1}& 1.175 & {\tt Ub2}& 0.951 \nl
{\tt Bb3}& 1.056 & {\tt Ha2}& 0.942 & {\tt Mb2}& 1.176 & {\tt Ub3}& 1.003 \nl
{\tt Bc1}& 0.781 & {\tt Ha3}& 1,053 & {\tt Mb3}& 1.013 & {\tt Uc1}& 0.936 \nl
{\tt Bc2}& 0.889 & {\tt Hb1}& 0.291 & {\tt Mc1}& 0.787 & {\tt Uc2}& 1.195 \nl
{\tt Bc3}& 0.816 & {\tt Hb2}& 0.311 & {\tt Mc2}& 0.797 & {\tt Uc3}& 0.910 \nl
{\tt Bd1}& 0.866 & {\tt Hb3}& 0.281 & {\tt Mc3}& 0.825 & {\tt Xa1}& 0.859 \nl
{\tt Bd2}& 0.762 & {\tt Hc1}& 0.853 & {\tt Md1}& 0.820 & {\tt Xa2}& 0.907 \nl
{\tt Bd3}& 0.829 & {\tt Hc2}& 0.993 & {\tt Md2}& 0.864 & {\tt Xa3}& 1.007 \nl
{\tt Be1}& 0.905 & {\tt Hc3}& 0.993 & {\tt Md3}& 0.869 & {\tt Xb1}& 0.959 \nl
{\tt Bf1}& 0.885 & {\tt Ka1}& 0.760 & {\tt Na1}& 0.684 & {\tt Xb2}& 0.966 \nl
{\tt Ca1}& 0.910 & {\tt Ka2}& 0.763 & {\tt Nb1}& 0.709 & {\tt Xb3}& 0.858 \nl
{\tt Ca2}& 1.094 & {\tt Ka3}& 0.808 & {\tt Nc1}& 0.949 & {\tt Xc1}& 1.280 \nl
{\tt Ca3}& 1.156 & {\tt Kb1}& 0.791 & {\tt Nd1}& 0.999 & {\tt Xc2}& 0.985 \nl
{\tt Cb1}& 0.880 & {\tt Kb2}& 0.759 & {\tt Ne1}& 0.664 & {\tt Xc3}& 1.236 \nl
{\tt Cb2}& 1.057 & {\tt Kb3}& 0.754 & {\tt Nf1}& 0.673 & {\tt Xd1}& 1.017 \nl
{\tt Cb3}& 1.101 & {\tt Kc1}& 0.571 & {\tt Pa1}& 0.768 & {\tt Xd2}& 1.175 \nl
{\tt Cc1}& 1.198 & {\tt Kc2}& 0.547 & {\tt Pa2}& 0.822 & {\tt Xd3}& 1.072 \nl
{\tt Cc2}& 0.893 & {\tt Kc3}& 0.629 & {\tt Pa3}& 0.884 & {\tt Xe1}& 0.854 \nl
{\tt Cc3}& 0.888 & {\tt Kd1}& 0.606 & {\tt Sa1}& 0.903 & {\tt Xf1}& 0.984 \nl
{\tt Cd1}& 0.939 & {\tt Kd2}& 0.566 & {\tt Sa2}& 1.013 & {\tt Xg1}& 0.433 \nl
{\tt Cd2}& 0.853 & {\tt Kd3}& 0.528 & {\tt Sa3}& 1.019 & {\tt Xg2}& 0.447 \nl
{\tt Cd3}& 0.964 & {\tt Ke1}& 0.606 & {\tt Sb1}& 0.961 & {\tt Xg3}& 0.474 \nl
\enddata
\end{deluxetable}

%%%%%%%%%%%%%%%%%%%%%%%%%%%%%%%%%%%%%%%%%%%%%%%%%%%%%%%%%%%%%%%%%%%%%%%%%%%%%%%
%
%                   REFERENCES

\clearpage
\begin{center}
Figure Captions
\end{center}

\figcaption{Left panels: parameter phase space for the database. Models
are indicated by dots. The number next to each dot indicates the 
number of simulations. Top panel $\Omega_0-\lambda_0$ phase space;
middle panel: $\Omega_0-H_0$ phase space; bottom panel: $\Omega_0-\sigma_8$
phase space. Right panels: same as left panels, but for a subregion
of the parameter phase space. Top panel: $H_0=0.65\,\rm km\,s^{-1}Mpc^{-1}$ 
models only; middle and bottom panels: $\lambda_0=0$ models only}

\figcaption{Sample FORTRAN program which reads a file from the database}

\figcaption{Relationships between the various values of
$\sigma_8$. Each dot represents one simulation. The dashed lines
indicate the equality between the values of $\sigma_8$ plotted.}

\figcaption{Two-point correlation function $\xi$ versus
separation in Mpc for models with $\Omega_0=0.2$ (solid curves).
The values of the parameters are indicated in each panel. Panels
with several curves show $\xi$ for models with various values of
$\sigma_8$. These values are indicated in the same order as the
curves, from top to bottom. The dashed line shows the power law
given by equation~(56). The thick dash indicates the location
of the softening scale $\epsilon=300\,\rm kpc$.}

\figcaption{Same as Fig.~4, except for models with $\Omega_0=0.3$ and 0.35.}

\figcaption{Same as Fig.~4, except for models with $\Omega_0=0.5$ and 0.7.}

\figcaption{Same as Fig.~4, except for models with $\Omega_0=1$.}


\clearpage

\begin{thebibliography}{}

\bibitem{}
Albretch, A., \& Steinhardt, P. 1982, Phys.Rev.Lett., 48, 1220

\bibitem{}
Babul, A., Weinberg, D. H., Dekel, A., \& Ostriker, J. P. 1994,
ApJ, 427, 1

\bibitem{}
Bahcall, N. A. 1999, preprint (astro-ph/9901076)

\bibitem{}
Bahcall, N. A., Cen, R., \& Gramann, M. 1993, ApJ, 408, L77

\bibitem{}
Bahcall, N. A., \& Fan, X. 1998, ApJ, 504, 1

\bibitem{}
Bahcall, N. A., Fan, X., \& Cen, R. 1997, ApJ, 485, 53

\bibitem{}
Bardeen, J. M., Bond, J. R., Kaiser, N., \& Szalay, A. S. 1986,
ApJ, 304, 15

\bibitem{}
Barlett, J. G., \& Silk, J. 1993, ApJ, 407, L45

\bibitem{}
Barnes, J., \& Efstathiou, G. 1987, ApJ, 319, 575

\bibitem{}
Barnes, J. E., \& Hut, P. 1986, Nature, 324, 446

\bibitem{}
Barnes, J. E., \& Hut, P. 1989, ApJ Suppl., 70, 389

\bibitem{}
Bertschinger, E. 1985a, ApJ Suppl., 58, 1

\bibitem{}
Bertschinger, E. 1985b, ApJ Suppl., 58, 39

\bibitem{}
Bouchet, F. R., \& Hernquist, L. 1992, ApJ, 400, 25

\bibitem{}
Briel, U. G., Henry, J. P., \& Boringer, H. 1992, A\&A, 259, L31

\bibitem{}
Bucher, M., Goldhaber, A. S., \& Turok, N. 1995, Phys.Rev.D, 52, 3314

\bibitem{}
Bunn, E. F., \& White, M. 1997, ApJ, 480, 6

\bibitem{}
Caldwell, R. R., Dave, R., \& Steinhardt, P. J. 1998, Phys.Rev.Lett., 80, 1582

%\bibitem{}
%Carlberg, R. G. 1988a, ApJ, 324, 664

%\bibitem{}
%Carlberg, R. G. 1988b, ApJ, 332, 26

\bibitem{}
Carlberg, R. G. 1994, ApJ, 433, 468

\bibitem{}
Carlberg, R. G., \& Couchman, H. M. P. 1989, ApJ, 340, 47

\bibitem{}
Carlberg, R. G., Yee, H. K. C., Ellington, E., Abraham, R.,
Gravel, P., Morris, S., \& Pritchet, C. J. 1996, ApJ, 462, 32

\bibitem{}
Centrella, J., \& Melott, A. L. 1983, Nature, 305, 196

\bibitem{}
Chaboyer, B. 1998, preprint (astro-ph/9808200)

\bibitem{}
Chaboyer, B., Demarque, P., Kernan, P. J., \& Krauss, L. M.  1998, 
ApJ, 494, 96

\bibitem{}
Charlton, J. C., \& Turner, M. S. 1987, ApJ, 313, 495

\bibitem{}
Coleman, S. 1988, Nucl.Phys.B, 307, 867

\bibitem{}
Colombi, S., Bouchet, F. R., \& Hernquist, L. 1996, ApJ, 465, 14

\bibitem{}
Copi, C., Schramm, D. N., \& Turner, M. S. 1995, Science, 267, 192

\bibitem{}
Couchman, H. M. P. 1991, ApJ, 368, L23

\bibitem{}
Davis, M., Efstathiou, G., Frenk, C. S., \& White, S. D. M. 1985,
ApJ, 292, 371

\bibitem{}
Efstathiou, G., 1995, MNRAS, 274, L73
  
\bibitem{}
Efstathiou, G., Davis, M., Frenk, C. S., \& White, S. D. M. 1985,
ApJ Suppl., 57, 241

\bibitem{}
Efstathiou, G., \& Eastwood, J. W. 1981, MNRAS, 194, 503

\bibitem{}
Evrard, A. E. 1986, ApJ, 310, 1

\bibitem{}
Evrard, A. E. 1987, ApJ, 316, 36

\bibitem{}
Fillmore, J. A., \& Goldreich, P. 1984a, ApJ, 281, 1

\bibitem{}
Fillmore, J. A., \& Goldreich, P. 1984b, ApJ, 281, 9

\bibitem{}
Freedman, W. L. 1998, Proc.Nat.Acad.Sci., 95, 2

\bibitem{}
Frenk, C. S., White, S. D. M., Davis. M., \& Efstathiou, G. 1988, 
ApJ, 327, 507

\bibitem{}
Fry, J. N. 1985, Phys.Lett.B, 158, 211

\bibitem{}
Fry, J. N., Melott, A. L., \& Shandarin, S. F. 1992, ApJ, 393, 431

\bibitem{}
Fukushige, T., Ito, T., Makino. J., Ebisuzaki, T., Sugimoto, D.,
\& Umemura, M. 1991, PASJ, 43, 841

\bibitem{}
Garnevich, P. M. et al. 1998, ApJ, 493, L53

\bibitem{}
Garriga, J., Tanaka, T., \& Vilenkin, A. 1998, preprint (astro-ph/9803268)

\bibitem{}
Gott, J. R., Gunn, J. E., Schramm, D. N., \& Tinsley, B. M. 1974,
ApJ, 194, 543

\bibitem{}
Gramann, M. 1988, MNRAS, 234, 569

\bibitem{}
Gramann, M., Cen, R., \& Bahcall, N. A. 1993, ApJ, 419, 440

\bibitem{}
Gross, M. A. K., Somerville, R. S., Primack, J. R., Holtzman, J., \& Klypin, A. 1998, MNRAS, 301, 81

\bibitem{}
Guth, A. 1981, Phys.Rev.D, 23, 347

\bibitem{}
Hale-Sutton, D., Fong, R., Metcalfe, N., \& Shanks, T. 1989, MNRAS, 237, 569

\bibitem{}
Hamana, T. 1998, private communication

\bibitem{}
Hawking, S. W. 1983, in Proc. 183 Shelter Island Conf. on
Quantum Field Theory and the Fundamental Problems of Physics, ed.
R. Jackiw et al. (Cambridge: MIT Press)

\bibitem{}
Hawking, S. W. 1984, Phys.Lett. B, 134, 403

\bibitem{}
Hernquist, L., Bouchet, F. R., \& Suto, Y. 1991, ApJ Suppl., 75, 231

\bibitem{}
Hockney, R. W., \& Eastwood, J. W. 1981, Computer Simulation
Using Particles (New York: McGraw-Hill)

\bibitem{}
Hu, W., \& Sugiyama, N. 1996, ApJ, 471, 542

\bibitem{}
Jimenez, R. 1998, preprint (astro-ph/9810311)

\bibitem{}
Jimenez, R., Thejll, P., J\o rgensen, U. G., MacDonald, J., \& Pagel, B. 1996,
MNRAS, 282, 926

\bibitem{}
Kaiser, N. 1984, ApJ, 284, L9

\bibitem{}
Klypin, A., \& Holtzman, J. 1997, preprint (astro-ph/9712217)

\bibitem{}
Klypin, A., Nolthenius, R., \& Primack, J. 1997, ApJ, 474, 533

\bibitem{}
Klypin, A., \& Shandarin, S. F. 1983, MNRAS, 204, 891

\bibitem{}
Kolb, E. W., \& Turner, M. S. 1990, The Early Universe 
(New York: Addison-Wesley)

\bibitem{}
Kravtsov, A. V., Klypin, A. A., \& Khokhlov, A. M. 1997, ApJ Suppl, 111, 73

\bibitem{}
Krauss, L. M. 1998, preprint (hep-ph/9807376)

\bibitem{}
Krauss, L. M., \& Kernan, P..J. 1995, Phys.Lett.B, 347, 347

\bibitem{}
Lahav, O., Lilje, P. B., Primack, J. R., \& Rees, M. J. 1991, MNRAS, 251, 128

\bibitem{}
Liddle, A. R., \& Lyth, D. 1993, Phys.Rep., 231, 1

\bibitem{}
Lin, H., Kirshner, R. P., Shectman, S. A., Landy, S. D.,
Oemler, A., Tucker, D. L., \& Schechter, P. L. 1996, ApJ, 471, 617

\bibitem{}
Linde, A. D. 1982, Phys.Lett B., 108, 289

\bibitem{}
Linde, A. D. 1986, Phys.Lett B., 175, 395

\bibitem{}
Linde, A. D. 1987, Phys.Scr., T15, 169

\bibitem{}
Linde, A. D. 1988, Phys.Lett B., 202, 194

\bibitem{}
Linde, A. D. 1995, Phys.Lett B., 351, 99

\bibitem{}
Linde, A. D., \& Mezhlumian, A. 1995, Phys.Rev.D, 52, 6789

\bibitem{}
Martel, H. 1990, Ph.D. Thesis (Cornell University)

\bibitem{}
Martel, H. 1991a, ApJ, 366, 353

\bibitem{}
Martel, H. 1991b, ApJ, 377, 7

\bibitem{}
Martel, H. 1995, ApJ, 445, 537

\bibitem{}
Martel, H., \& Shapiro, P. R. 1998, MNRAS, 297, 467 

\bibitem{}
Martel, H., Shapiro, P. R., Valinia, A., \& Villumsen, J. V. 1994
in Dark Matter, eds. S. S. Holt and C. L. Bennett, AIP
Conference Proceedings 336, 441

\bibitem{}
Martel, H., Shapiro, P. R., \& Weinberg, S. 1998, ApJ, 492, 29

\bibitem{}
Melott, A. L. 1986, Phys.Rev.Lett, 56, 1992

\bibitem{}
Melott, A. L., \& Shandarin, S. F. 1993, ApJ, 410, 469

\bibitem{}
Miller, R. H., 1983, ApJ, 270, 390

\bibitem{}
Moore, B., Katz, N., \& Lake, G. 1996, ApJ, 457, 455

\bibitem{}
Moutarde, F., Alimi, J.-M., Bouchet, F. R., Pellat, R., \& Ramani, A. 1991,
ApJ, 382, 377

\bibitem{}
Navarro, J. F., Frenk, C. S., \& White, S. M. D. 1997, ApJ, 490, 493

\bibitem{}
Park, C., Gott, J. R., Melott, A. L., \& Karachentsev, I. D. 1992,
ApJ, 387, 1

\bibitem{}
Peacock, J. A., \& Dodds, S. J. 1994, MNRAS, 267, 1020

\bibitem{}
Peebles, P. J. E. 1980, The Large-Scale Structure of the Universe
(Princeton: Princeton University Press)

\bibitem{}
Peebles, P. J. E. 1993, Physical Cosmology 
(Princeton: Princeton University Press)

\bibitem{}
Pen, U.-L. 1995, ApJ Suppl., 100, 269 

\bibitem{}
Premadi, P., Martel, H., \& Matzner, R. 1998, ApJ, 493, 10

\bibitem{}
Press, W. H., \& Schechter, P. 1974, ApJ, 187, 425

\bibitem{}
Perlmutter, S. et al. 1998, ApJ, in press (astro-ph/9812133)

\bibitem{}
Ratra, B., \& Peebles, P. J. E. 1994, ApJ, 432, L5

\bibitem{}
Shandarin, S. F. 1980, Astrophizika, 16, 769

\bibitem{}
Shapiro, P. R., Struck-Marcell, C., \& Melott, A. L. 1983, ApJ, 275, 413

\bibitem{}
Silveira \& Waga 1994, Phys.Rev.D., 50, 4890

\bibitem{}
Smoot, G. F. et al. 1992, ApJ, 396, L1

\bibitem{}
Tegmark, M., Eisenstein, D. J., Hu, W., \& Kron, R., G. 1998, submitted
to ApJ (astro-ph/9803117)

\bibitem{}
Thomas, P. A. et al. 1998, MNRAS, 296, 1061

\bibitem{}
Viana, P. T. P., \& Liddle, A. R. 1996, MNRAS, 281, 323

\bibitem{}
Vilenkin, A. 1995, Phys.Rev.Lett., 74, 846

\bibitem{}
Villumsen, J. V. 1989, ApJ Suppl., 71, 407

%\bibitem{}
%Walker, T. P., Steigman, G., Schramm, D. N., Olive, K. A., \&
%Kang, H.-S. 1991, ApJ, 376, 51.

\bibitem{}
Weinberg, S. 1996, in Critical Dialogues in Cosmology, ed. N. Turok
(Singapore: World Scientific), 1

\bibitem{}
West, M. J., Oemler, A., \& Dekel, A. 1989, ApJ, 346, 539

\bibitem{}
West, M. J., Villumsen, J. V., \& Dekel, A. 1991, ApJ, 369, 287

\bibitem{}
White, M. 1998, ApJ, 506, 495

\bibitem{}
White, S. D. M., Frenk, C. S., \& Davis, M. 1983,
ApJ, 274, L1

\bibitem{}
White, S. D. M., Frenk, C. S., Davis, M., \& Efstathiou, G. 1987a,
ApJ, 313, 505

\bibitem{}
White, S. D. M., Davis, M., \& Efstathiou, G., \& Frenk, C. S. 1987b,
Nature, 330, 451

\bibitem{}
White, S. D. M., Navarro, J. F., Evrard, A. E., \& Frenk, C. S. 1993, Nature,
366, 429

\bibitem{}
Yamamoto, K., Sasaki, M., \& Tanaka, T. 1995, ApJ, 455, 412

\bibitem{}
Yess, C., \& Shandarin, S. F. 1996, ApJ, 465, 2

\bibitem{}
Zel'dovich, Ya. B. 1970, A\&A, 5, 84

\end{thebibliography}
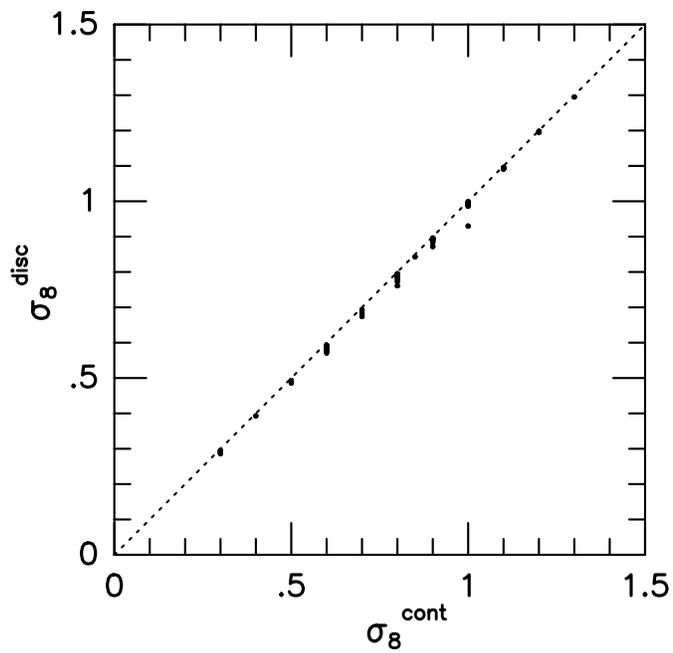
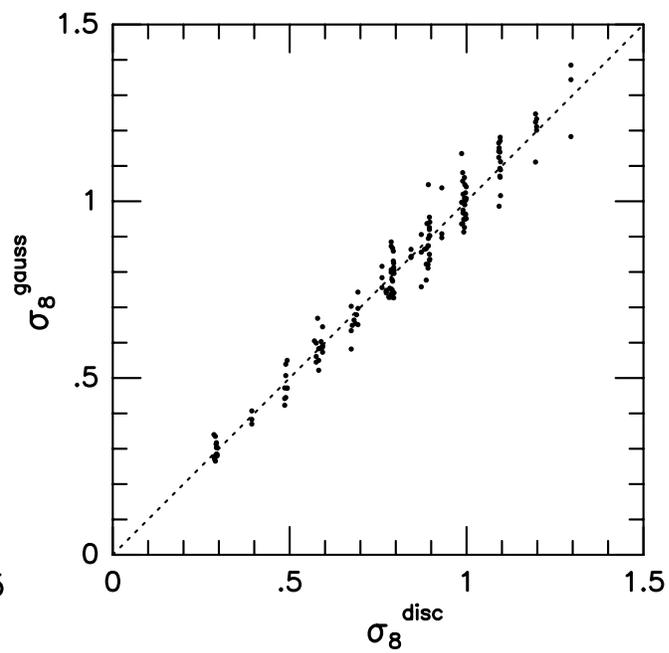
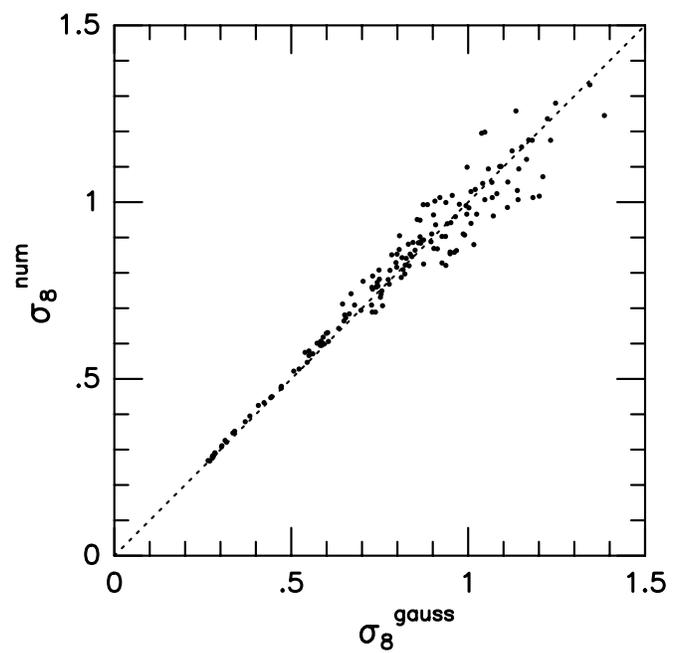
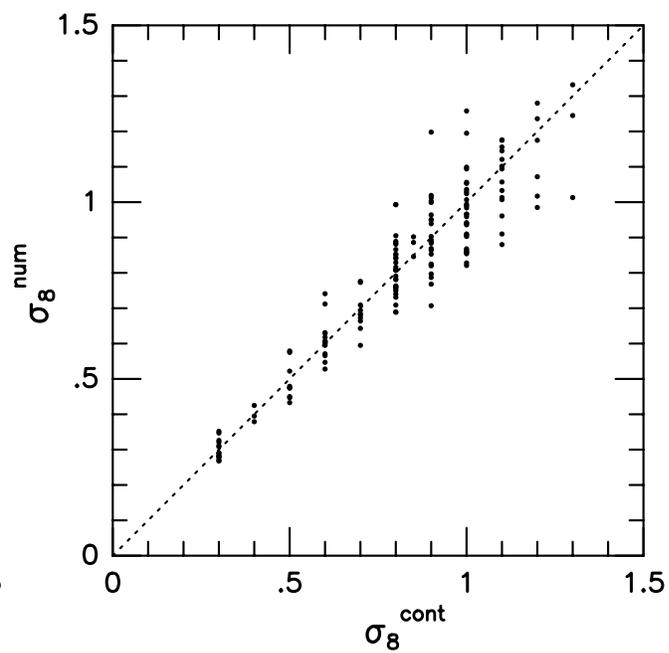
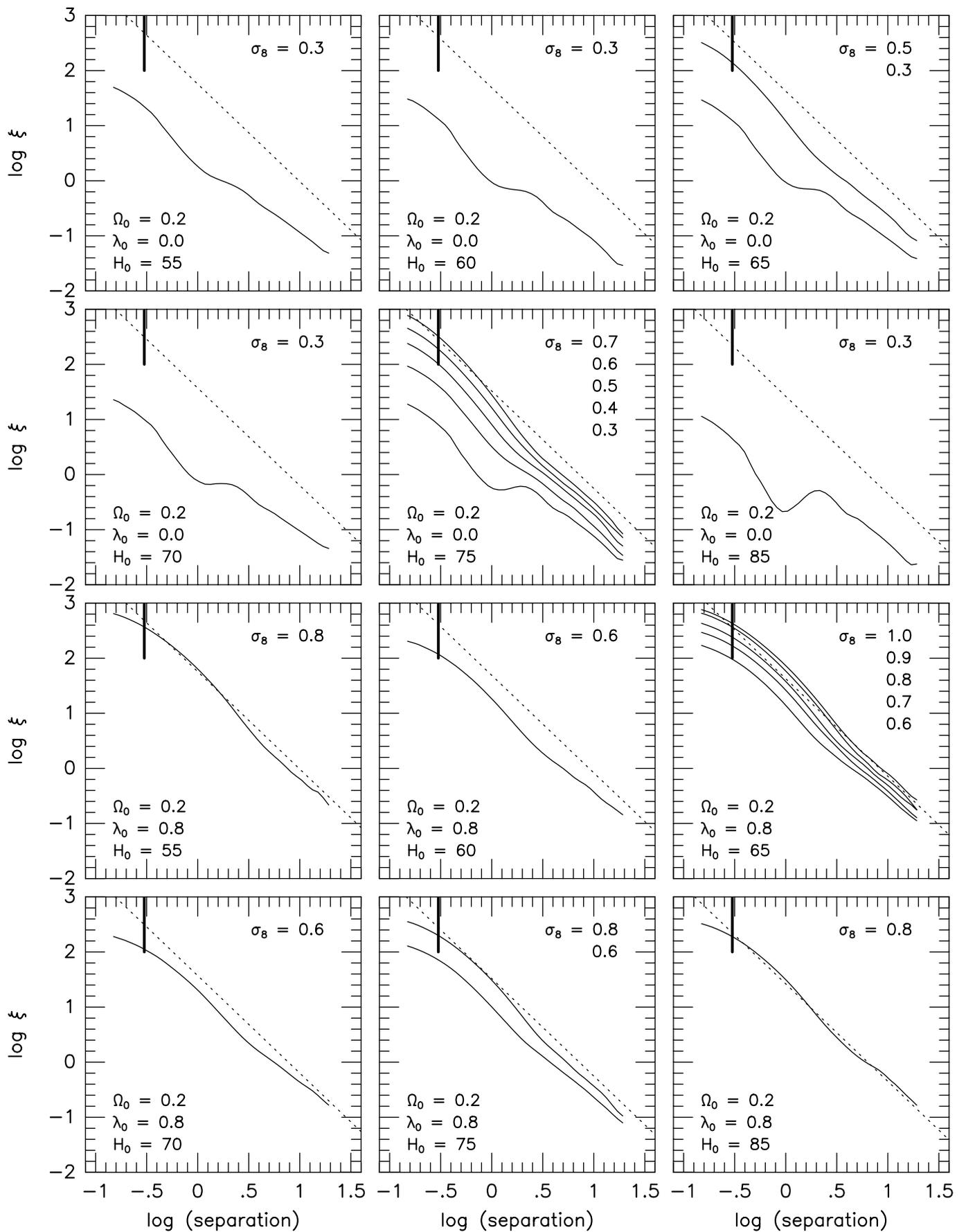
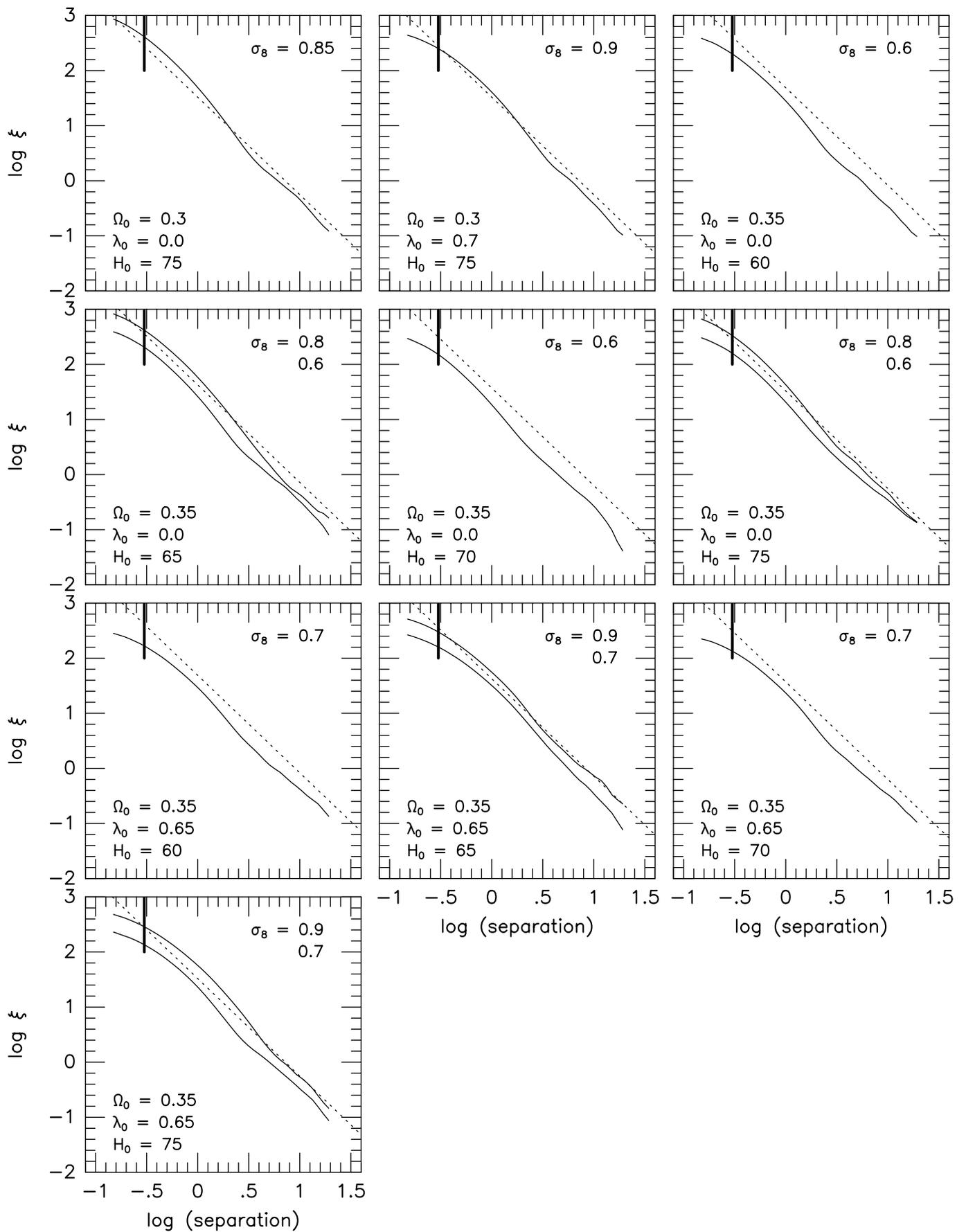
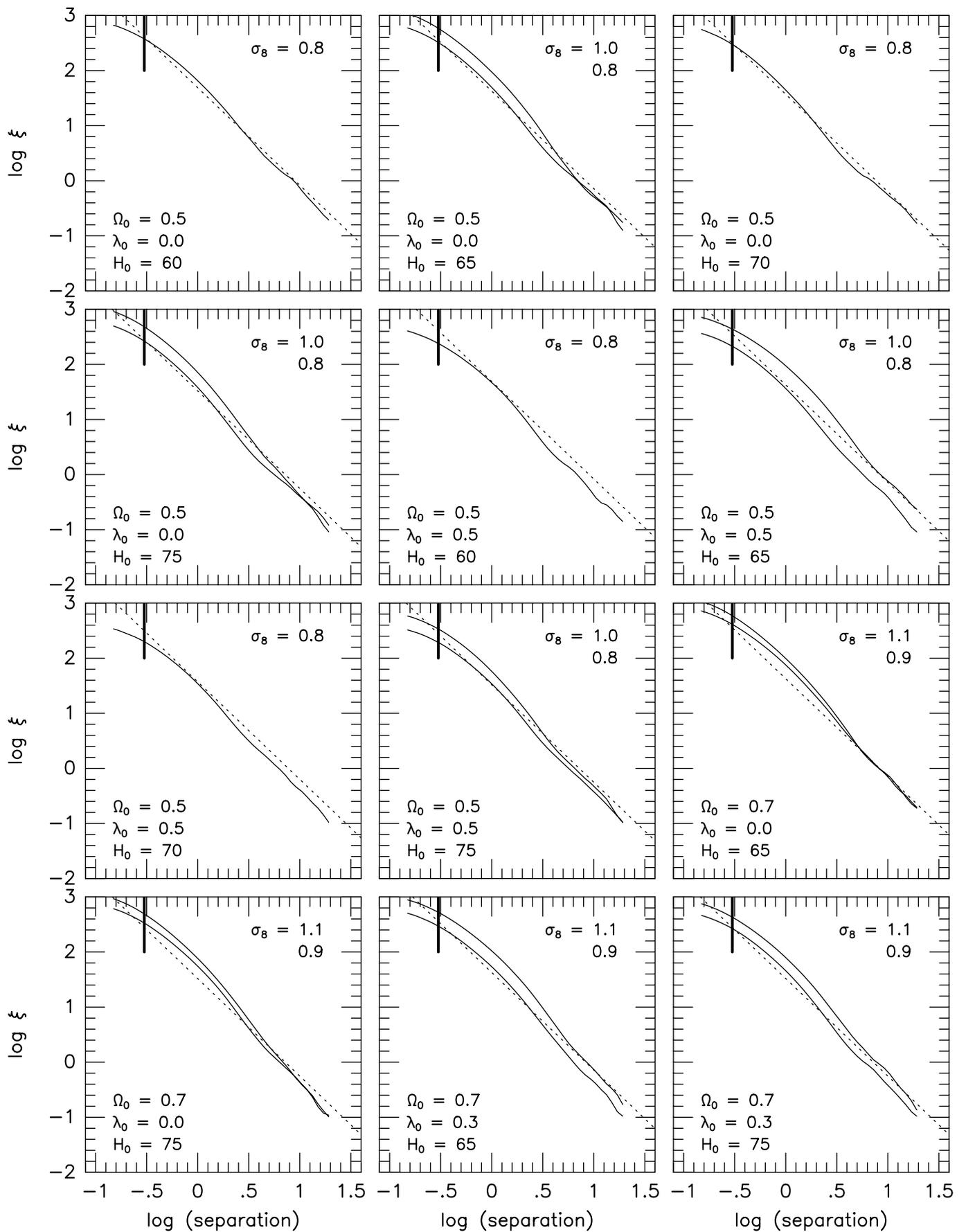
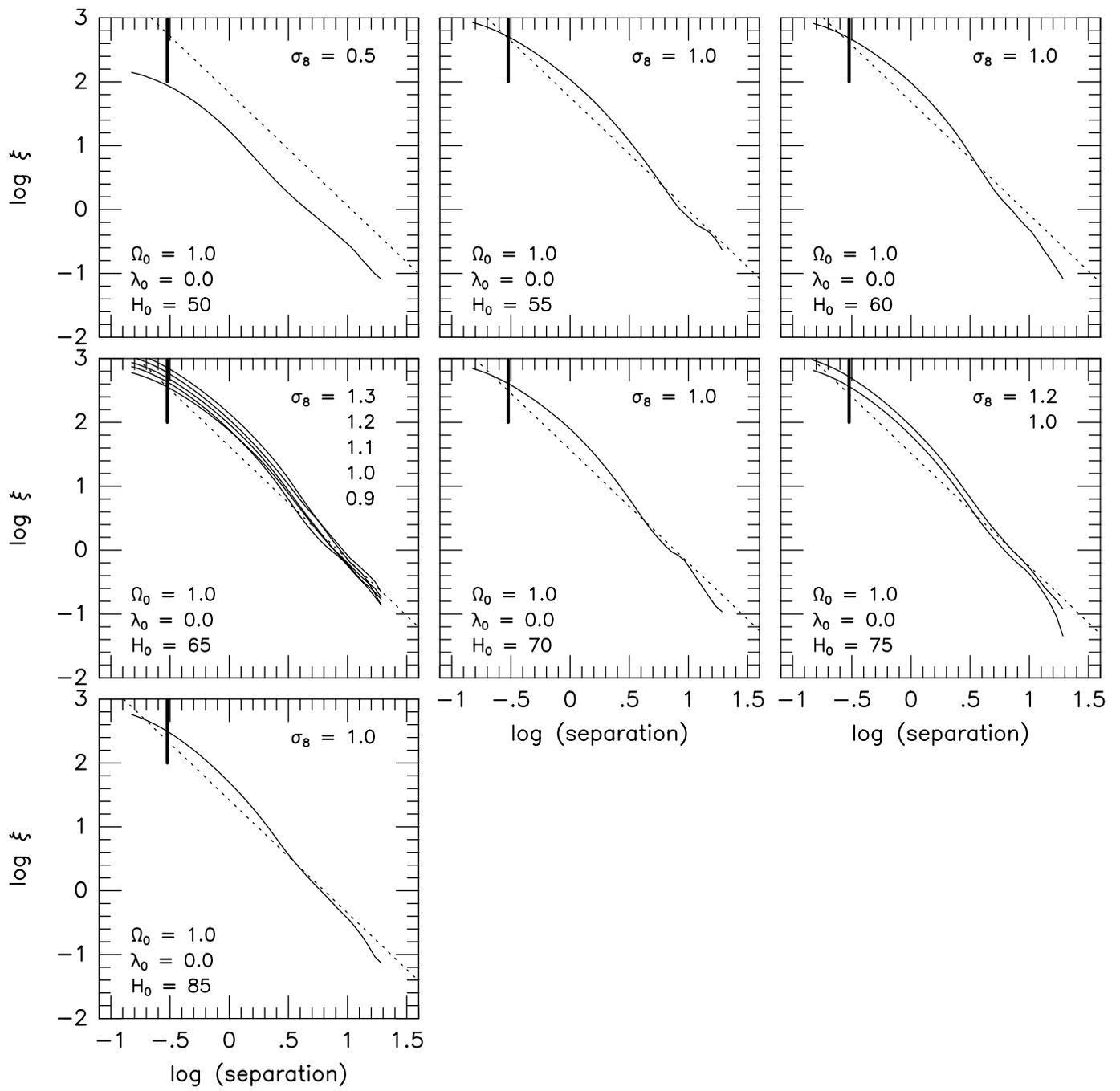
\end{document}